\documentclass[conference]{IEEEtran}
\usepackage{lipsum}  
\usepackage{xspace}
\usepackage{comment}
\usepackage{xcolor,soul}
\usepackage[skip=2pt]{caption}
\usepackage{subcaption}
\usepackage{graphicx}
\usepackage{xspace}
\usepackage{tabularx}
\usepackage[utf8]{inputenc}
\usepackage[T1]{fontenc}
\usepackage{booktabs} 
\usepackage{diagbox}
\usepackage{hyperref}
\hypersetup{
    breaklinks=true,
    colorlinks=true,
    linkcolor=blue,
    filecolor=magenta,      
    urlcolor=cyan,
    pdftitle={ML-based Behavioral Malware Detection is Far from a Solved Problem}
    }

\usepackage{float}
\usepackage{enumitem}
\usepackage{adjustbox}
\usepackage{multirow}
\usepackage[T1]{fontenc}
\usepackage{tikz}
\usepackage{fancyvrb}

\usepackage{amsmath}
\usepackage[nameinlink,capitalize,noabbrev]{cleveref}
\usepackage{mathtools}
\usepackage{amsthm}
\usepackage{pifont}
\usepackage{tcolorbox}

\usepackage[nocompress]{cite}

\PassOptionsToPackage{hyphens}{url}
\PassOptionsToPackage{breaklinks=true}{hyperref}
\usepackage{xurl}

\definecolor{lightgray}{gray}{0.85}

\newcommand\resultno[1]{{\texttt{EX#1}}}

\newcommand\amdelta{{\texttt{AM\_Delta}}\xspace}

\newcommand\roblox{{\texttt{Roblox}}\xspace}

\newcommand\freefootnote[1]{
    \vspace{0.2cm}
    \hrule
    \vspace{-0.2cm}
    \begingroup
    \renewcommand\thefootnote{}\footnote{#1}
    \addtocounter{footnote}{-1}
    \endgroup
}

\newcommand{\arxivtext}{This work has been accepted for publication in the IEEE Conference on Secure and Trustworthy Machine Learning (SaTML). The final version will be available on IEEE Xplore.}

\newcommand*\circled[1]{\tikz[baseline=(char.base)]{\node[shape=circle,draw,inner sep=1pt, draw=blue, text=black] (char) {\textbf{#1}};}}

\newcommand{\ndtopic}[1]{\vspace{0.03in}\noindent\textbf{#1}}
\newcommand{\topic}[1]{\vspace{0.03in}\noindent\textbf{#1.}}

\newcommand{\seq}{\,{=}\,} 

\newcommand{\ssbsb}[2]{\textbf{\texttt{SB#1->SB#2}}\xspace}
\newcommand{\ssbep}[1]{\textbf{\texttt{SB#1->EP}}\xspace}
\newcommand{\sepep}{\textbf{\texttt{EP->EP}}\xspace}

\newcommand{\RNum}[1]{\uppercase\expandafter{\romannumeral #1\relax}}

\newcommand\bnumber[1]{(\textbf{#1})}
\newcommand\ie{\emph{i.e.}, }
\newcommand\eg{\emph{e.g.}, }

\newcommand\maldy{\texttt{NGR}\xspace}
\newcommand\neurlux{\texttt{HYB}\xspace}
\newcommand\nebula{\texttt{ATT}\xspace}

\newcommand{\modification}[1]{#1}

\newcommand\takeaways[1]{\begin{tcolorbox}[boxsep=1pt,left=3pt,right=3pt,top=3pt,bottom=3pt,width=\columnwidth,colframe=darkgray]
\textbf{\modification{Takeaways:}} #1
\end{tcolorbox}
}

\crefformat{section}{\S#2#1#3} 
\crefformat{subsection}{\S#2#1#3}
\crefformat{subsubsection}{\S#2#1#3}
\Crefformat{appendix}{\S#2#1#3}

\begin{document}


\title{ML-Based Behavioral Malware Detection Is Far From a Solved Problem}


\author{
    \IEEEauthorblockN{Yigitcan Kaya} 
    \IEEEauthorblockA{\textit{UC Santa Barbara}\\
    yigitcan@ucsb.edu}
    \and
    \IEEEauthorblockN{Yizheng Chen} 
    \IEEEauthorblockA{\textit{Univ. of Maryland, College Park}\\
    yzchen@umd.edu}
    \and
    \IEEEauthorblockN{Marcus Botacin} 
    \IEEEauthorblockA{\textit{Texas A\&M University}\\
    botacin@tamu.edu}
    \and
    \IEEEauthorblockN{Shoumik Saha} 
    \IEEEauthorblockA{\textit{Univ. of Maryland, College Park}\\
    smksaha@umd.edu}
    \and
    \IEEEauthorblockN{Fabio Pierazzi} 
    \IEEEauthorblockA{\textit{University College London}\\
    f.pierazzi@ucl.ac.uk}
    \and
    \IEEEauthorblockN{Lorenzo Cavallaro} 
    \IEEEauthorblockA{\textit{University College London}\\
    l.cavallaro@ucl.ac.uk}
    \and
    \IEEEauthorblockN{David Wagner} 
    \IEEEauthorblockA{\textit{UC Berkeley}\\
    daw@cs.berkeley.edu}
    \and
    \IEEEauthorblockN{Tudor Dumitra\c{s}} 
    \IEEEauthorblockA{\textit{Univ. of Maryland, College Park}\\
    tudor@umd.edu}
}

\maketitle

\begin{abstract}
Malware detection is a ubiquitous application of Machine Learning (ML) in security.
In behavioral malware analysis, the detector relies on features extracted from program execution traces.
The research literature has focused on detectors trained with features collected from sandbox environments and evaluated on samples also analyzed in a sandbox.
However, in deployment, a malware detector at endpoint hosts often must rely on traces captured from endpoint hosts, not from a sandbox.
Thus, there is a gap between the literature and real-world needs.

We present the first measurement study of the performance of ML-based malware detectors at real-world endpoints.
Leveraging a dataset of sandbox traces and a dataset of in-the-wild program traces, we evaluate two scenarios: \bnumber{i} an endpoint detector trained on sandbox traces (convenient and easy to train), and \bnumber{ii} an endpoint detector trained on endpoint traces (more challenging to train, since we need to collect telemetry data).
We discover a wide gap between the performance as measured using prior evaluation methods in the literature---over 90\%---vs. expected performance in endpoint detection---about 20\% (scenario \bnumber{i}) to 50\% (scenario \bnumber{ii}).
We characterize the ML challenges that arise in this domain and contribute to this gap, including label noise, distribution shift, and spurious features.
Moreover, we show several techniques that achieve 5--30\% relative performance improvements over the baselines.
Our evidence suggests that applying detectors trained on sandbox data to endpoint detection is challenging. 
The most promising direction is training detectors directly on endpoint data, which marks a departure from current practice.
To promote progress, we will facilitate researchers to perform realistic detector evaluations against our real-world dataset.

\end{abstract}
\section{Introduction}
Detection of malware threats is crucial for governments, enterprises, and end users as there are significant (and growing) financial and safety harms of malware infections~\cite{reddick_2022}, which has created a \$7 Billion industry in 2022 with many players~\cite{imarcreport}.
Malware detection appears to be remarkably effective: 
industry-standard evaluations of commercial anti-malware products~\cite{avcomparisons}, and prior research on malware detection using machine learning (ML) methods~\cite{raff2018malware,karbab2019maldy}, routinely report that over 99\% of malware samples in a standard corpus can be detected, with very few false positives.

\freefootnote{\arxivtext}

\begin{table*}[t]
\centering
\begin{tabular}{l|l|l|l}
\toprule
Training data &Test data &Method &TPR @ 1\%FPR\\
\midrule
Sandbox traces &Sandbox traces &Standard classifier &$\sim$95\% \cite{karbab2019maldy,jindal2019neurlux,trizna2023nebula}\\
Sandbox traces &Endpoint traces &Standard classifier &$\sim$17\% (ours)\\
Endpoint traces &Endpoint traces &Standard classifier &$\sim$49\% (ours)\\
Sandbox traces &Endpoint traces &Training set resampling + invariant learning &$\sim$22\% (ours)\\
Endpoint traces &Endpoint traces &Soft labels + invariant learning &$\sim$52\% (ours)\\
\bottomrule
\end{tabular}
\caption{Performance of ML-based behavioral malware detection under different settings.}
\label{table:overview}
\vspace{-0.35cm}
\end{table*}

Malware detection is often performed at endpoints, where an endpoint security solution~\cite{quarantine} monitors host devices to detect threats.
In practice, these solutions employ a chain of techniques~\cite{vmray}, including static analysis and dynamic analysis. 
Static methods, such as blocklists and signatures, operate without executing the program.
As static methods can readily be bypassed via obfuscation or polymorphism~\cite{moser2007limits,song2007infeasibility}, dynamic analysis has become a standard offering~\cite{kasperskyrealtime}.

Dynamic analysis relies on observing the execution behaviors of a sample to detect whether it displays malicious behaviors.
The typical paradigm for obtaining a behavioral detector is to detonate (execute) a large set of known samples (malware and benign) in a controlled environment (a \emph{sandbox}), collect their execution traces, and learn to separate malicious and benign behaviors. 
Several ML models have been effective in this task~\cite{mariconti2016mamadroid,karbab2019maldy,jindal2019neurlux,trizna2023nebula}.
Unfortunately, endpoint detection models cannot use sandbox traces to decide whether a sample under observation is malicious, as detonation cannot be done in real time, despite advances such as on-premise sandboxes~\cite{onpremsandbox}.
Consequently, these models must rely on traces observed at real-world hosts to detect malware.

Research has outlined several challenges associated with this task.
First, program behaviors are environment-sensitive, and, especially for malware, two traces of the same sample from two different hosts often have little overlap~\cite{avllazagaj2021malware}.
As a result, an ML model trained on a trace captured on one host might fail to generalize to others.
Second, techniques frequently employed to evade sandbox analysis~\cite{balzarotti2010efficient,galloro2022systematical} can mislead models trained on sandbox traces into learning spurious features that do not capture actual malware behaviors.
Although these challenges have been previously articulated in the research literature~\cite{barecloud,kirat2015malgene,jindal2019neurlux,avllazagaj2021malware}, we lack an understanding of how they manifest and can be addressed.

Our work conducts the first quantitative study into the efficacy of ML-based methods for endpoint detection to demystify these challenges.
We specifically focus on two scenarios: \bnumber{i} an endpoint detector trained using only sandbox traces and \bnumber{ii} an endpoint detector trained using real-world endpoint traces.
Existing techniques can conveniently gather large datasets of sandbox traces, making \bnumber{i} a practical and accessible option.
On the other hand, collecting endpoint traces is attainable mainly by vendors who receive telemetry data from the wild, as simulating such data at scale in controlled environments (such as sandboxes) is still an open problem~\cite{schreuders2017security,miramirkhani2017spotless,mills2020investigating}.
This makes \bnumber{ii} a less accessible but likely a more effective option.
Consequently, both scenarios are relevant in practice, which motivates us to study their challenges.

We evaluate three ML approaches~\cite{karbab2019maldy,jindal2019neurlux,trizna2023nebula} that reach 95\% true-positive rate (TPR) at 1\% false-positive rate (FPR) when evaluated on sandbox traces.
We use a dataset from Avllazagaj et al.~\cite{avllazagaj2021malware} containing around 1M endpoint traces from 26K samples, recorded on real-world hosts by an anti-malware vendor.
The malicious samples in this dataset were undetected by the vendor's defenses at that time and caused real-world infections.
Consequently, this data reflects the realistic threats that behavioral detectors must combat.
Additionally, we collect a dataset of traces (contemporaneous to our endpoint data) from two sandboxes to realistically study scenario \bnumber{i}.

Our initial finding is that ML methods perform poorly in both scenarios \bnumber{i} ($\sim$17\% TPR) and \bnumber{ii} ($\sim$49\% TPR), in contrast with their excellent performance on sandbox traces (see Table~\ref{table:overview}).
We study the low performance in \bnumber{i} from the lens of distribution shift, a problem that plagues ML in many applications~\cite{koh2021wilds}.
%
Similarly, the security community has articulated the challenge of concept drift, where the data distribution shifts over time~\cite{jordaney2017transcend}.
We attempt to address the differences in data distribution between sandbox vs. endpoint traces using classification with rejection (an existing tool against concept drift~\cite{jordaney2017transcend}) and found that it does not improve the performance enough.
Interestingly, endpoint traces of benign samples are also rejected as ``drifting'' in high proportions, whereas in prior work concept drift has mostly affected malware~\cite{barbero2022transcending}.

Investigating why existing methods perform poorly, we first discovered that endpoint detectors are applied to a different distribution of samples than considered in prior work: they are typically applied only to the hardest-to-classify samples.
Past research has trained and evaluated detectors on a corpus of samples from repositories~\cite{karbab2019maldy,jindal2019neurlux,dambra2023decoding} such as MalwareBazaar~\cite{bazaar}.
However, endpoint malware detection systems employ a pipeline where basic methods (\eg static signatures) attempt to categorize samples first, and ML-based classifiers are applied only to samples that remain unresolved~\cite{vmray}.
%
%
Consequently, in the wild, endpoint classifiers are only applied to samples that tend to be harder to classify than an average sample in a standard corpus.
Prior research has overlooked this factor, and we find that it causes a significant drop in performance: it reduces the performance of classifiers trained on sandbox traces from 95\% (for samples from a standard corpus) to $\sim$60\% TPR (when we adjust the distribution of samples to take into account earlier stages in the pipeline).
This shows that prior evaluations have vastly overestimated the effectiveness of behavioral detectors in the wild.

Second, we study the impact of variable program behaviors across different environments.
We discovered large differences between a sample's sandbox trace and its endpoint traces.
Sandbox traces lack diversity: collecting multiple traces by running a group of related samples, \eg from the same malware family, in a sandbox produces very similar traces.
This introduces spurious features that do not generalize to other environments.
In contrast, endpoint traces are diverse, which hurts performance by making a model's predictions on different traces of the same sample inconsistent.
Moreover, we provide the first evidence that sandbox-evading malware (40--80\% of all malware~\cite{galloro2022systematical}) biases a model in scenario \bnumber{i} to classify very short traces (often an indication of evasion~\cite{kuchler2021does}) as malware.
This correlation is spurious as it is absent in endpoint traces; thus, it causes prior evaluations using sandbox traces to overestimate the accuracy of endpoint classifiers.

Characterizing these challenges allows us to explore avenues for performance improvements.
To improve the performance on the distribution of difficult-to-classify samples, we use soft labeling (effective against label noise~\cite{lukasik2020does}) and more accurate distributions for training.
Against variable behaviors, we employ a technique popular in other areas of ML: invariant learning~\cite{zhao2019learning}.
In particular, we train our models to make consistent predictions on different traces of the same sample, forcing them to learn robust, environment-independent features.
Together, these strategies lead to moderate gains, from 17\% to 22\% TPR in scenario \bnumber{i} and from 49\% to 52\% in scenario \bnumber{ii} (30\% and 5\% relative improvement, respectively).

We believe our results should serve as a call to action for the community.
Prior research has suggested that it is possible to achieve over 95\% TPR, which might leave the impression that progress is saturated and there is not much room for improvement in behavioral malware detection using ML.
We show that the reality is different: the problem is not solved, the actual performance on malware in the wild is far worse, and there are major unsolved challenges and significant room for further improvement.
Moreover, ML methods that have shown success on standard benchmarks, such as group robustness~\cite{sagawa2020investigation}, struggle due to the complexities of this domain.
We plan to stimulate progress on this problem by releasing our sandbox dataset and metadata and offering a pipeline that allows researchers to evaluate their behavioral malware detectors against our real-world endpoint data.
Our public website (\href{https://malwaredetectioninthewild.github.io/}{https://malwaredetectioninthewild.github.io/}) includes the details on this evaluation pipeline and data release.

\topic{Contributions}
\textbf{(I)} We measure the performance discrepancy of ML-based malware detectors between sandbox-only and endpoint settings. 
\textbf{(II)} We characterize the ML challenges, such as distribution shifts, behind this discrepancy.
\textbf{(III)} We explore ML methods to improve the endpoint performance.

\section{Background and Related Work}

\topic{Dynamic Malware Analysis}
Most work in dynamic analysis focuses on analyzing the behaviors of a sample detonated in controlled environments, such as sandboxes~\cite{willems2007toward,bayer2009view}.
As dynamic analysis has become a staple, malware has started including checks that suppress malicious activities if the environment is fingerprinted as a sandbox, known as evasive malware~\cite{lindorfer2011detecting}.
Although many strategies have been developed to prevent fingerprinting~\cite{liu2022enhancing}, this is still an ongoing arms race~\cite{miramirkhani2017spotless}.
Even the methods that analyze samples in \emph{bare metal} environments~\cite{balzarotti2010efficient,barecloud} struggle to prevent evasion~\cite{miramirkhani2017spotless}.
To our knowledge, we have taken the first step toward measuring the implications of evasion for ML-based detectors at endpoints.
A recent large-scale measurement study over variable program behaviors in the wild has found that a given malware sample can behave significantly differently across time and in different real-world hosts~\cite{avllazagaj2021malware}. 
Our work connects to this line of work as we are interested in quantifying the impact of these challenges, such as evasion or variability, on a malware detector deployed for endpoint detection at hosts in the wild.

\topic{ML for Malware Detection}
ML-based methods are widely used in research and practice with static~\cite{raff2018malware} and behavioral features~\cite{huang2016mtnet}.
In behavioral detection, methods that treat a program's execution trace as a text document (a sequence of tokens) and adapt existing ML architectures have shown promise~\cite{karbab2019maldy,jindal2019neurlux,trizna2023nebula}.
Most of these methods are trained and tested on the traces from a pre-configured sandbox for samples collected from public repositories (see~\cite {awesomemalware} for the popular ones).
We aim to understand the implications of these practices and the efficacy of popular ML approaches for detecting malware with behavioral features in real-world hosts.

\topic{Distribution Shifts in ML}
Distribution shift occurs when a model's training and testing distributions have significant differences, which hurts the performance~\cite{koh2021wilds}.
For example, as new malware variants emerge and old ones disappear over time, the performance of a detector that has not been kept up-to-date will deteriorate, known as concept drift~\cite{pendlebury2019tesseract}.
An effective way to deal with concept drift is deploying drift detectors to reject samples that would have been misclassified and training on them later to update the model~\cite{jordaney2017transcend,barbero2022transcending,chen2023continuous}.
We deploy a drift detector to characterize how the distribution shifts in our problem differ from concept drift.
The ML community has also proposed other ideas to tackle distribution shifts in different contexts, such as domain invariant features~\cite{ganin2015unsupervised, 2013gongconnecting}, distributionally robust optimization~\cite{sagawa2020investigation}, or continuous learning~\cite{chen2023continuous}.
We adapt some of these ideas, such as invariant learning, to assess their benefits in our problem.

\topic{Limitations of ML for Security}
Research suggests that popular ML methods have many pitfalls when applied to security tasks~\cite{arp2020and}.
For example, in malware detection, the success of many methods has been overestimated due to impossible time splits of training and testing sets~\cite{pendlebury2019tesseract}.
Another line of work focuses on how ML methods successful in lab-only evaluations break down in the real world due to distribution shifts, \eg in the context of network anomaly detection~\cite{jacobs2022ai}, website fingerprinting~\cite{cherubin2022online} or malware detection~\cite{das2019sok,aghakhani2020malware}.
In these contexts, models are known to learn spurious features, such as specific IP addresses~\cite{jacobs2022ai}, that are artifacts of the experimental setup and do not apply to realistic settings.
We investigate the limitations of ML methods specifically in behavioral malware detection when they are trained and evaluated in controlled settings (\eg using sandboxes) but deployed in the wild.
\section{Endpoint Malware Detection}
\label{sec:malware_detection}

\topic{Terminology}
A \emph{sample} is an executable program identified by its unique SHA-256 hash.
A \emph{trace} is a sequence of behavioral actions (such as file accesses or process creations) performed by a sample when it is executed in a computing environment.
We refer to an in-the-wild endpoint environment as a \emph{host}.
Our work focuses on Windows hosts and executable samples.
In our endpoint dataset, the anti-malware system in each host recorded the traces of samples executed by the host that were not determined to be benign or malicious at an earlier stage of the detection pipeline.
Each sample can have multiple traces recorded at multiple hosts at different times.
A sandbox (\textbf{SB}) is a synthetic environment that provides tools for analyzing samples and recording their traces with controlled executions.
Our sandbox dataset contains traces collected from two commercial sandboxes: Tencent HABO~\cite{habo} and VirusTotal~\cite{virustotal}, which we will refer to as SB1 and SB2, respectively.

\topic{Notation}
We denote an individual sample in our dataset as $P_i$ and its ground truth label as $y_i$, where $y_i\!=\!0$ and $y_i\!=\!1$ indicate a benign and malware sample, respectively.
If $P_i$ is a malware sample, it is also tagged with a family label $s_i$ that is useful for grouping samples with similar characteristics.
A trace of $P_i$ is $x_i^j$, where $j$ denotes the execution environment, specifically, $j\!=\!0$ refers to the endpoint hosts and $j\!\in\!\{1,2\}$ refers to the sandboxes.
As there are multiple endpoint traces per sample, we refer to the $k$-th one as $x_{i,k}^0$, and the endpoint traces of a sample are enumerated by their timestamps. 
The timestamp of its earliest observed trace---$x_{i,0}^0$---marks the first time $P_i$ was first seen in the wild, and $\mathbf{x}^0_{i, (t<h)}$ is the set of all endpoint traces recorded within $h$ hours of this.

An ML-based detection model takes a trace $x_i$ of $P_i$ as its input and aims to infer $y_i$.
We split a model $f$ into two parts: an encoder $\text{enc}$ and a classifier $g$.
The encoder produces a vector embedding $z_i = \text{enc}(x_i)$ of the input trace and $f(x_i) = g(z_i)$ is the model's predicted probability (\emph{score}) that $P_i$ is malware.
The predicted label $\hat{y_i}$ is $\hat{y_i}\!=\!1$ if $f(x_i)\!\geq\!\tau$, or $\hat{y_i}\!=\!0$, otherwise.
Here, $\tau$ is a tunable threshold, where a higher $\tau$ reduces false positives in exchange for fewer true positives. 
In \cref{ssec:success_metrics}, we discuss the tuning of $\tau$ in more detail.

\begin{figure}[hbt]
\centering
\includegraphics[width=\columnwidth]{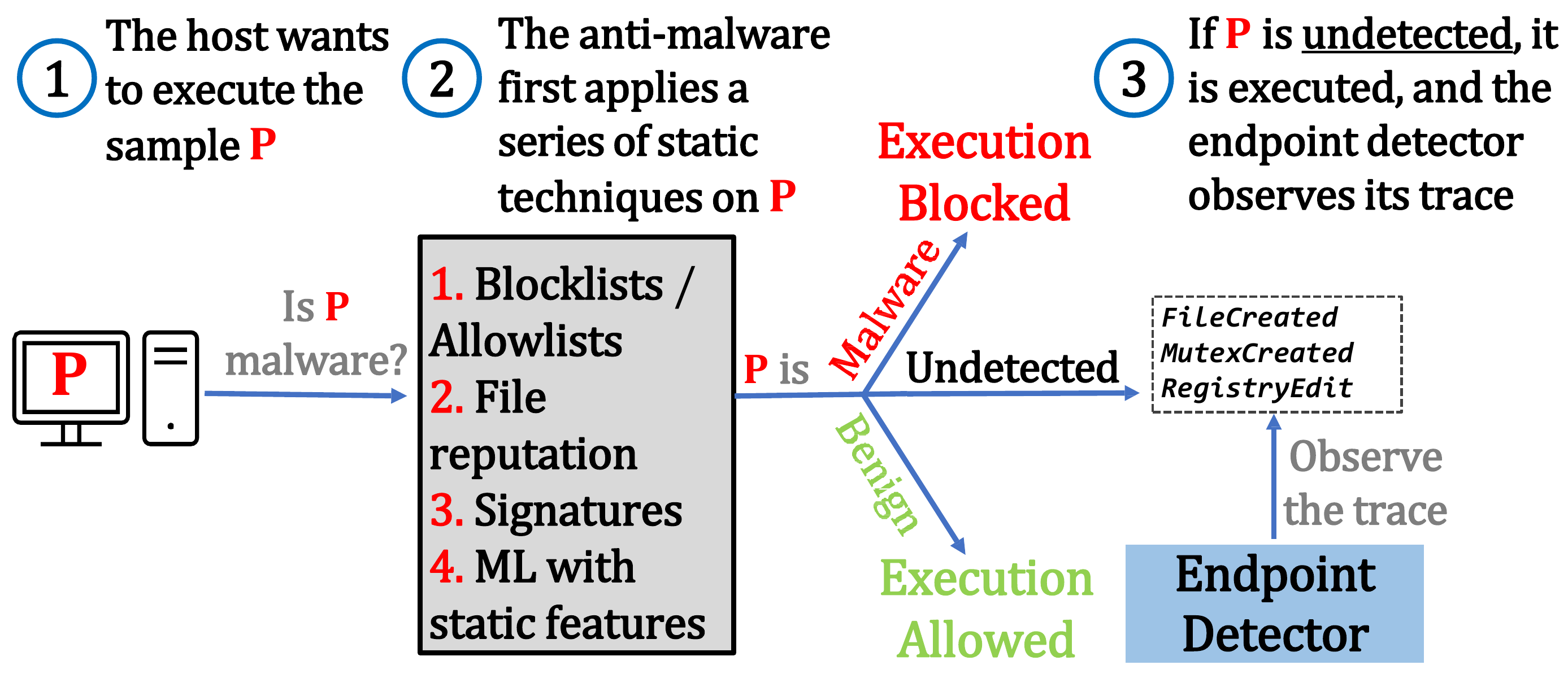}
  \caption{\modification{An overview of endpoint malware detection.}}
  \label{fig:mal_workflows}
  \vspace{-0.18in}
\end{figure}

\subsection{\modification{Malware Detection Pipeline}}

\label{ssec:detection_workflows}

Figure~\ref{fig:mal_workflows} provides an overview of the pipeline for malware detection at endpoints.
In practice, vendors deploy a chain of techniques~\cite{avcomparisons,vmray} to detect if a sample $P$ encountered by a host (\eg from an e-mail attachment) is malware---step \circled{2} and \circled{3}.
If one technique in the pipeline fails to classify the sample as malware or benign conclusively, the next one is invoked.
First, it deploys high-precision methods (blocklists and signatures), then less precise static ML approaches~\cite{anderson2018ember,raff2018malware}, and finally, dynamic and behavioral analysis. 
We focus on ML-based behavioral malware detectors deployed in step \circled{3}.
Static techniques operate without executing the sample and are deployed first as they are more efficient and secure---step \circled{2}.
This is sufficient for classifying most existing samples~\cite{dambra2023decoding} either as malware ($P$ is prevented from executing) or benign ($P$ is allowed to execute).
If $P$ remains \emph{undetected} after this step, it will be executed if the host demands it, \eg by clicking on the attachment.
If $P$ is malware, executing it leads to a real-world compromise---an \emph{infection}.
An endpoint detector is the last line of defense that operates at the host in real-time to identify whether $P$ is malware by consuming the trace resulting from this execution---step \circled{3}.
If $P$ is detected as malware during this step, anti-malware products often apply remedies such as quarantining, deleting files, or alerting~\cite{quarantine}.

We identified two issues regarding this pipeline in past research.
First, prior work constructed evaluation sets for behavioral malware detectors without considering their position in an end-to-end pipeline.
As a result, the challenges in performing behavioral detection on samples that bypassed static methods are unknown.
Second, endpoint detectors must classify the sample based on an endpoint trace observed at the end host.
It is not feasible to first execute the sample in a sandbox and then classify it based on the sandbox trace, as running in a sandbox (whether in the cloud~\cite{cloudsandbox} or on-premise~\cite{onpremsandbox}) introduces unacceptable delays.
Prior work has only evaluated ML detectors on sandbox traces, but in deployment, the detectors will be applied to endpoint traces, so their actual performance has not been accurately evaluated.
We tackle these issues by performing realistic evaluations using our endpoint dataset that contains only the traces of undetected samples---which reached the step \circled{3} in the wild.

\topic{\modification{Assumptions and Limitations}}
Security vendors such as VMRay~\cite{vmray}, Palo Alto Networks~\cite{paloalto}, and Kaspersky~\cite{kasperskybeh} use multi-tiered solutions that chain static and dynamic analysis in their anti-malware products.
We do not have access to these proprietary pipelines.
Instead, we consider a streamlined one in Figure~\ref{fig:mal_workflows} based on public documentation by these vendors.
Such a pipeline should ideally be evaluated end-to-end, starting with a large pool of samples from the wild, where each component sees only the samples that previous components cannot classify.
We are unable to perform such an evaluation as we do not have raw binaries (to extract static features) of our endpoint samples (we could find less than 10\% of samples from public sources, mostly malware).
Prior malware detection datasets (\eg SOREL~\cite{harang2020sorel20m}) only share vectorized static features and not the binaries, preventing the simulation of behavioral components.
However, we can still realistically isolate and evaluate behavioral ML methods at the pipeline's last stage as our endpoint dataset contains samples that bypassed a real-world product's static components.
We recommend future work to collect datasets (with static and dynamic features) that allow the evaluation of the whole pipeline end-to-end or each component according to its position.

Moreover, a pipeline's components constantly evolve in practice.
For example, a sample from a new malware family might initially bypass static methods.
Eventually, this family can be caught at step \circled{2} with new static signatures or updates to static ML models.
Evaluating this requires a dataset with a large temporal span, whereas our endpoint data covers only six months (insufficient for significant changes~\cite{barbero2022transcending}).
Consequently, we perform our evaluation with fixed behavioral ML models that are trained at a particular timestamp and tested on future samples that bypass static methods.

\subsection{Model Training and Evaluation Scenarios}
\label{ssec:characterizing_scenarios}

There are two main phases for deploying an ML-based endpoint detector: \emph{training} and \emph{testing}.
In the training phase, the vendor collects a set of samples.
These samples are often collected from large-scale repositories, either public, such as MalwareBazaar~\cite{bazaar}, or private~\cite{ugarte2019close}.
Most prior work in this domain has followed this practice~\cite{jindal2019neurlux,miller2016reviewer,karbab2019maldy,dambra2023decoding,kuchler2021does}.
The vendor can rely on a public platform, such as VirusTotal~\cite{virustotal}, to label these samples or employ malware analysts in more advanced cases~\cite{yong2021inside}.
Most commonly, the collected samples are executed in sandboxes that can conveniently produce large-scale datasets of traces for training.
In contrast, collecting a large-scale dataset of endpoint traces from hosts requires a telemetry infrastructure, a permissive user agreement, and a large enough user base---conditions that commercial anti-malware vendors can meet but researchers typically cannot.
The collected traces are used to train an ML model that can detect malware from runtime behaviors.
During the test phase, the model is deployed locally at each host.
Training and testing phases can also be interwoven as vendors continuously update their models on new data~\cite{chen2023continuous}.

Table~\ref{table:detection_scenarios} summarizes the detection scenarios, each defined along four dimensions, considered in our work.
The \textsc{Samples} dimensions indicate the step of the pipeline training or testing samples originate from: \circled{1}---a broad distribution of all programs hosts encounter---or \circled{3}---a distribution of programs that remained undetected after \circled{2}.
The \textsc{Env.} (short for environment) dimensions encode where the traces are collected---from a sandbox or endpoint hosts.

\begin{table}[hbt]
\normalsize
\centering
\begin{adjustbox}{max width=0.95\columnwidth}
\begin{tabular}{c|cc||cc}
\toprule
\multirow{2}{1.75cm}{\textbf{Deployment Scenario}}
 & \multicolumn{2}{c||}{\textbf{Training Phase}} & \multicolumn{2}{c}{\textbf{Testing Phase}} \\
& Samples & Env. & Samples & Env. \\
\midrule
\ssbsb{}{} & Step \circled{1} & Sandbox & Step \circled{1} & Sandbox \\
\ssbep{}{} & Step \circled{1} & Sandbox & Step \circled{3} & Endpoint \\
\sepep & Step \circled{3} & Endpoint & Step \circled{3} & Endpoint \\
\bottomrule
\end{tabular}
\end{adjustbox}
\caption{Scenarios for behavioral malware detection.}
\label{table:detection_scenarios}
\vspace{-0.15in}
\end{table}

\ssbsb{}{} is the scenario most prior work in our domain considers: both training and testing samples are collected from sample repositories and corpora, which represent the distribution of all samples seen in the wild, and these samples are executed in a sandbox~\cite{karbab2019maldy,jindal2019neurlux,miller2016reviewer,kuchler2021does}.
It is already known (and also corroborated by our results) that detectors excel (90\%+ TPR) in this scenario~\cite{dambra2023decoding}.

In \ssbep{} and \sepep scenarios, detectors are deployed at \circled{3} in Figure~\ref{fig:mal_workflows}, \ie on the endpoint traces of undetected samples.
In \ssbep{}, the detector is trained only on sandbox-based data (same as \ssbsb{}{}) but deployed for endpoint detection.
In \sepep, the model is trained on the real-world endpoint traces of undetected samples.
This departs from the norm in published research, as most prior detectors have relied on sandbox traces for training.
We study this scenario to assess the success of a model trained explicitly for endpoint detection.
These scenarios represent a trade-off between practicality (where \ssbep{} is favored) and detection performance (where \sepep is favored).

\subsection{Success Metrics for Endpoint Detection}
\label{ssec:success_metrics}
A conventional detector aims to detect as many malware samples as possible, \ie to maximize the true-positive rate (TPR), while minimizing the number of benign samples misclassified, \ie the false-positive rate (FPR).
The defender can control the TPR and FPR by tuning the detection threshold $\tau$; a higher $\tau$ implies a lower TPR and FPR.
Following prior work~\cite{arp2014drebin}, we evaluate our models regarding their TPR @1\% FPR on the test sets (reported as TPR).
Additionally, we report the area under the TPR-FPR curve, \ie the receiver operating characteristic curve, to gauge a model's overall ability to separate malware and benign classes (reported as AUC).
Note that a uniform random predictor, regardless of the class proportions in the test set, would achieve 1\% TPR @1\% FPR and 50\% AUC.

Measuring TPR and FPR in \ssbsb{}{} is straightforward as there is a one-to-one mapping from samples to traces.
However, in endpoint detection (\ssbep{} and \sepep), there is a one-to-many mapping as multiple hosts may execute the same sample at different times, and these resulting traces often have high variability~\cite{avllazagaj2021malware}.
For example, a few malware samples have over 400 traces, each corresponding to a real-world infection (see Figure~\ref{fig:trace_stats_numbers}).
To obtain a one-to-one mapping, we consider ``a representative trace'' by randomly selecting one endpoint trace of each sample to create an EP test set on which we measure a model's TPR.
We repeat this process 100 times and report the average TPR over these runs as the final TPR.
This gives us a fair and unambiguous way to compare performance across scenarios.
We refine this metric further based on practical considerations.
67\% of traces of an average malware sample (52\% for benign) are seen within 24 hours of its first execution at a host.
As this ratio drops rapidly after the first day (see Figure~\ref{fig:trace_stats_days}) and malware samples die within days~\cite{yong2021inside}, we only randomly select from the first-day traces---$\mathbf{x}^0_{i, (t<24)}$---unless stated otherwise.

\section{Technical Setup}

\subsection{Datasets}
\label{ssec:datasets}

\begin{table}[hbt]
\large
\centering
\begin{adjustbox}{max width=0.75\columnwidth}
\begin{tabular}{lrr|rr}
\toprule
& \multicolumn{2}{c|}{\textbf{Training}} & \multicolumn{2}{c}{\textbf{Testing}} \\
 \textbf{Dataset} & Mal. & Ben. & Mal. & Ben. \\
\midrule
EP & 0.5K & 16.5K & 0.4K & 8.1K \\
EP (Traces) & 19.4K & 531.6K & 9.7K & 412.5K \\
\midrule
SB1 & 46.4K & 16.7K & 31.0K & 16.4K \\
SB2 & 34.5K & 7.8K & 18.1K & 7.3K \\
SB1 $\cap$ SB2  & 9.9K & 5.3K & 8.9K & 5.5K \\
\bottomrule
\end{tabular}
\end{adjustbox}
\caption{A summary of the samples in our datasets.}
\label{table:dataset_stats}
\vspace{-0.3cm}
\end{table}

\topic{Endpoint (EP) Dataset}
Provided by Avllazagaj et al.~\cite{avllazagaj2021malware}, we use a dataset of program traces recorded on real Windows hosts of a commercial anti-malware vendor from over 100 countries between January and July 2018.
The vendor did not know whether the samples were benign or malicious at the time of execution.
The samples were executed by the users, who interacted with them naturally, and the vendor's behavioral component recorded the traces in a last-ditch effort to discover unknown threats. 
We have reprocessed this dataset for our problem and relabeled it by querying VirusTotal~\cite{virustotal} for more accurate labels.
Our processed dataset contains around 1M execution traces from 900 malware and 25K benign samples.
Each trace includes high-level actions (file, registry, process, and mutex actions) of a sample that is executed until its termination.
To our knowledge, this is the only dataset that allows us to evaluate endpoint detection at scale.

\topic{Sandbox (SB) Dataset}
A realistic study of \ssbep{} requires a sandbox dataset contemporaneous to our EP dataset.
Because the sandbox dataset will serve as the training set and the EP dataset as the test set, we must collect samples seen in or before 2018 to respect causality~\cite{pendlebury2019tesseract}.
Moreover, we cannot detonate samples in a sandbox, as it is commonly done~\cite{karbab2019maldy,jindal2019neurlux,miller2016reviewer,kuchler2021does}, as behaviors of old programs today would differ from their original behaviors.
This is because malware stops functioning when, for example, its remote infrastructure dies~\cite{yong2021inside} or it starts behaving differently over time~\cite{avllazagaj2021malware}.
Our solution is collecting traces from VirusTotal~\cite{virustotal}, where third-party sandbox vendors publicly share behavior reports on samples.
To this end, we curate a list of SHA-256 hashes of Windows samples from popular public malware detection corpora, including EMBER~\cite{anderson2018ember}, and SOREL~\cite{harang2020sorel20m}, released in 2017 or 2018.
We then query VirusTotal with these hashes to collect their sandbox traces when available.
We discard traces that came more than six months after the sample was first seen to capture close-to-original behaviors.
This results in traces from two sandbox vendors for around 110K malware and 40K benign samples.
These vendors are well-known in their countries of origin (SB1 is from China, and SB2 is from the USA).
There is a class imbalance as vendors on VirusTotal are more inclined to share traces of malware samples.

\begin{table*}[t!]
\normalsize
\begin{adjustbox}{max width=\textwidth}
\begin{tabular}{l|rrr|rrr|rrr|rrr|rrr|rrr|rrr}
\toprule
\multirow{3}{*}{\shortstack[l]{{Sel.}\\{Test}\\{Set}}} & 
\multicolumn{9}{c|}{\textbf{\large Trained on SB1}} & 
\multicolumn{9}{c|}{\textbf{\large Trained on SB2}} & 
\multicolumn{3}{c}{\textbf{\large Trained on EP}} \\

& \multicolumn{3}{c|}{\normalsize \ssbsb{1}{1}} &  
\multicolumn{3}{c|}{\normalsize \ssbsb{1}{2}} & 
\multicolumn{3}{c|}{\normalsize \ssbep{1}} & 
\multicolumn{3}{c|}{\normalsize \ssbsb{2}{1}} & 
\multicolumn{3}{c|}{\normalsize \ssbsb{2}{2}} & 
\multicolumn{3}{c|}{\normalsize \ssbep{2}} &
\multicolumn{3}{c}{\normalsize \sepep} \\ 

 & \maldy & \neurlux & \nebula & \maldy & \neurlux & \nebula & \maldy & \neurlux & \nebula & \maldy & \neurlux & \nebula & \maldy & \neurlux & \nebula & \maldy & \neurlux & \nebula & \maldy & \neurlux & \nebula \\
SB1 & 
95.0 & 94.2 & 93.2 & 
60.7 & 53.7 & 56.7 & 
11.2 & 8.0 & 4.5 & 

48.5 & 31.8 & 32.7 & 
85.5 & 59.2 & 83.7 & 
14.2 & 7.2 & 12.3 & 

43.5 & 34.0 & 25.7 \\  

SB2 &  
94.0 & 93.1 & 92.2 & 
70.4 & 61.2 & 64.3 & 
13.9 & 7.1 & 5.5 & 

17.6 & 10.0 & 14.7 & 
92.6 & 91.4 & 90.8 & 
7.5 & 6.3 & 11.4 & 

43.4 & 37.6 & 26.4 \\  

EP &  

93.2 & 93.1 & 87.8 & 
63.5 & 48.0 & 52.9 & 
16.7 & 10.2 & 8.3 & 

34.5 & 16.6 & 15.7 & 
79.3 & 79.5 & 88.7 & 
\color{red}{\textbf{17.3}} & 10.9 & 13.1 & 

\color{red}{\textbf{49.5}} & 43.5 & 42.8 \\  

\bottomrule
\end{tabular}
\end{adjustbox}
\caption{The performance (TPR\%) of three ML approaches (\maldy, \neurlux, \nebula) in seven detection scenarios (based on Table~\ref{table:detection_scenarios}). The models are selected using the test set in each row, and the average TPR of the top 20 models is reported.}
\label{table:tpr_results}
\vspace{-0.15in}
\end{table*}

\topic{Trace Standardization}
Our traces from three sources (SB1, SB2, and EP) have different formats, contents, and conventions.
We first pre-process our sandbox traces to keep only the type of behaviors recorded in our EP traces, such as file, process, mutex, and registry key creations.
We then convert all traces to a single standardized format (see~\cref{app:standardization} for details).
As shown in~\cref{sec:sb_limitations}, the resulting sandbox traces allow us to train models with comparable performance to prior work ($\sim$95\% TPR and 99\% AUC) in the \ssbsb{}{} scenario.
%

\topic{Temporal Splitting and Labeling}
We split our datasets into two portions based on the timestamps of the samples (the first-seen dates) to ensure that the train and test samples are temporally disjoint, avoiding a common pitfall in prior work~\cite{pendlebury2019tesseract,arp2020and}.
Samples seen before April 1st, 2018, and their traces are in the \emph{Training} portion, and the samples seen after that are in the \emph{Testing} portion, on which we never train.
As labels of samples are known to change mildly over time~\cite{labeldynamics}, we make the best effort to assign the training samples historical labels that were available before April 1st, 2018.
We could find such historical labels for $\sim$24\% and 100\% of the samples in the EP and SB training portions, respectively.
For testing samples, we query VirusTotal to obtain the most recent detection reports and label samples detected by over five anti-malware engines as malware, following the advice in~\cite{labeldynamics}.
Further, we use AVClass2~\cite{sebastian2020avclass2} to assign family labels to our malware samples based on their detection reports.
We tag malware samples that did not receive a family label as \texttt{Generic} malware.
This methodology ensures that our evaluation is realistic and reflects the conditions the anti-malware vendor had to work under when our endpoint data was collected.

\subsection{Machine Learning Details}
\label{ssec:ml_details}
We experiment with three deep-learning-based approaches: \bnumber{i} a bag-of-n-grams-based model (\textbf{\maldy}), \bnumber{ii} a hybrid model that combines convolution and attention (\textbf{\neurlux)}, and \bnumber{iii} a self-attention-based sequence model (\textbf{\nebula)}.
These approaches cover a wide range of designs proposed by prior work in behavioral malware detection with ML, respectively, MalDy~\cite{karbab2019maldy}, Neurlux~\cite{jindal2019neurlux} and, most recently, Nebula~\cite{trizna2023nebula} (see~\cref{app:ml_details} for details on these models).
We address the class imbalance in our training sets by oversampling the underrepresented classes (malware in EP and benign in SB datasets).
Note that our goal is \emph{not} to develop a new architecture or feature processing routine but to evaluate the state-of-the-art ones at endpoint malware detection.
We make our ML improvements without changing the fundamentals of existing approaches.

\topic{Model Selection}
In each setting, we train a set of models with a hyper-parameter grid over model capacity (layer widths), learning rate, regularization (dropout), early-stopping, oversampling factor, and other setting-specific parameters, such as $\alpha$ in~\cref{sec:environment_shift}.
From this set, we report the average performance of the top-$N$ models, selected according to the metric of interest evaluated on a test set, \eg TPR on SB1, TPR on EP, or AUC on EP.
We opt for performing model selection using the test sets (instead of creating separate validation sets) because our EP dataset contains only a small number of malware samples.
We set $N\seq20$ so that we can still attribute our results to underlying learning approaches rather than having found a lucky set of hyper-parameters using a test set.
\section{Endpoint Detection Evaluations}
\label{sec:sb_limitations}

Table~\ref{table:tpr_results} presents the detection performance of ML models (based on the methodology in~\cref{ssec:ml_details}), using three different data sets (SB1, SB2, and EP) for training and testing, following the scenarios in Table~\ref{table:detection_scenarios}.
Here, for example, \ssbsb{1}{2} and \ssbep{1} refer to the scenarios where the models are trained on the SB1 training set and evaluated on the SB2 and EP testing sets, respectively.
See Table~\ref{table:app_auc_results} for the AUC results.

\subsection{Evaluating the \ssbep{} Scenario}
\label{ssec:eval_sbep}
We see wide performance gaps between \ssbsb{}{} and \ssbep{} scenarios.
The gap widens when models are trained and selected using the same sandbox, \eg 95.0\% vs. 11.2\% for \maldy trained and selected using SB1.
When models are selected using a different sandbox, the gap shrinks, \eg 94.0\% vs. 13.9\% for \maldy trained on SB1 and selected using SB2.
The gap is still substantial even when we select the models based on their EP performance: \eg 93.2\% vs. 16.7\% for \maldy trained on SB1 and selected using EP.
Although there is a gap between a model's performances on different sandboxes, \eg \ssbsb{1}{1} vs. \ssbsb{1}{2}, it is smaller than the \ssbep{} gap.
Overall, this gap persists regardless of the training sandbox or the model type, hinting at the limitations of sandbox-trained models for endpoint detection.

We observed that the gap widens when model training and selection use traces from the same sandbox.
In~\cref{app:model_selections}, we perform an experiment showing this is a generally valid observation.
We believe this is because a model that performs better on the traces from an unseen sandbox is less likely to have overfitted to features specific to the training sandbox.
We discuss such features in more detail in~\cref{sec:environment_shift}.
This implies that studies that rely on a single sandbox for training and evaluation in \ssbsb{}{}~\cite{miller2016reviewer,karbab2019maldy,kuchler2021does,trizna2023nebula} are at a higher risk of producing worse models for \ssbep{}.
Consequently, we recommend performing model selection in \ssbep{} using traces from a different, unseen sandbox.

Finally, we explore combining traces from two sandboxes (SB1 and SB2) for training.
However, this brings only minor improvements for \ssbep{} over training on traces from one sandbox.
When we use EP traces for model selection, training on both sandboxes achieves at most 17.5\% TPR on the EP test set for \maldy (11.7\% for \neurlux, and 13.9\% for \nebula).
In~\cref{sec:environment_shift}, we show that the diversity between the traces of a sample from two sandboxes is much lower than between its traces from two endpoint hosts.
This lack of diversity makes combining SB1 and SB2 traces ineffective for increasing the EP performance.

\topic{\modification{The Impact of Trace Standardization}}
The standardized trace format (see~\cref{ssec:datasets}) removes features from sandbox traces unavailable in endpoint traces.
In~\cref{app:standardization_impact}, to understand whether this might hurt the \ssbep{} performance, we experiment with training and testing models on unstandardized sandbox traces.
This shows that standardization moderately hurts the in-domain performance (test traces from the training sandbox) while improving generalization to other domains (traces from other sandboxes).
Consequently, we believe standardization helps with improving out-of-domain performance in \ssbep{}.

\topic{Applying Drift Detection}
We study the low performance in \ssbep{} through the lens of \emph{distribution shift}, which occurs if the process generating inputs, \emph{i.e.}, program traces, changes between the training and testing phases.
Concept drift is a highly-studied type of distribution shift in malware detection that stems from malware evolution~\cite{pendlebury2019tesseract}.
In past work on malware concept drift, the distribution shift builds up over time in concept drift, \eg over 50\% performance drop in 2 years~\cite{barbero2022transcending}.
In contrast, we find that in \ssbep{}, the shift manifests immediately after the model is deployed.

One way to combat malware concept drift is to use a drift detector (DD) to identify samples that might have drifted from the training distribution~\cite{jordaney2017transcend,barbero2022transcending,yang2021cade,chen2023continuous}.
The classifier then rejects such \emph{out-of-distribution} (OOD) samples while accepting the \emph{in-distribution} (ID) ones on which predictions are reliable~\cite{jordaney2017transcend,yang2021cade}.
We apply drift detection to quantify the distribution shift present in \ssbep{}.
We employ a recent technique ReAct~\cite{sun2021react} to assign an \emph{outlierness} score to each sample, among which the highest-scoring ones are rejected.
ReAct rectifies a model's penultimate layer activations, which calibrates the prediction probabilities of a model to be a better indicator of outlierness.
We selected ReAact due to its flexible approach, which can be applied to any neural network model.


We experiment with a \maldy model trained on the SB1 training set.
%
We tune ReAct to reject $K \in [0, 
15
]$\% of the traces in the SB1 test set (\ie the ID rejection rate) and separately measure for EP and SB2 test sets \bnumber{i} corresponding rejection rates on the malware and benign traces; and \bnumber{ii} the TPR on the accepted traces.
Table~\ref{table:rejection_tpr} presents the results.

Regarding \bnumber{i}, both EP and SB2 traces are rejected at much higher rates than the ID samples, \eg when $K\seq10$\%, ReAct rejects 53\% and 38\% of the malware traces in EP and SB2, respectively.
Malware traces in EP have the highest rejection rate, suggesting that they are the most \emph{drifted} from the malware traces in the SB1 training set.
That said, benign traces are also likely to be rejected, \eg when $K\seq10$\%, 39\% of the EP, and 37\% of the SB2 benign traces are rejected.
This is a critical distinction from past work on malware concept drift, where malware samples are rejected at much higher rates than benign samples~\cite{barbero2022transcending}.
The following sections aim to illuminate this phenomenon both qualitatively and quantitatively.

Regarding \bnumber{ii}, classification with rejection increases the performance for both EP and SB2 test sets.
However, this increase is less pronounced for EP, \eg gaining $\sim$10\% TPR (over the baselines in Table~\ref{table:tpr_results}) takes rejecting 67\% and 22\% of malware traces in EP and SB2, respectively.
This also distinguishes the distribution shift in \ssbep{} from concept drift; rejection rates must be high for meaningful performance gains due to the magnitude of the shift.

\begin{table}[hbt]
\normalsize
\begin{adjustbox}{max width=\columnwidth}
\begin{tabular}{c|rrr|rrr}
\toprule
\multirow{2}{*}{\shortstack[l]{($K$) ID \\ Rej\%}} & \multicolumn{3}{c|}{\normalsize \ssbep{1}} & \multicolumn{3}{c}{\normalsize  \ssbsb{1}{2}} \\
& \small{TPR} & Mal\% & Ben\% & \small{TPR} & Mal\% & Ben\%  \\
\midrule
5\% & 20.2\% & 31.1\% & 20.5\% & 79.2\% & 22.0\% & 21.5\% \\
10\% & 23.5\% & 53.2\% & 39.0\% & 89.4\% & 38.2\% & 36.5\% \\
15\% & 25.6\% & 67.7\% & 46.4\% & 88.3\% & 52.4\% & 53.6\%  \\

\bottomrule
\end{tabular}
\end{adjustbox}
\caption{The TPR of an \maldy model in \ssbsb{1}{2} and \ssbep{1} scenarios, with increasing ID rejection rates and the corresponding rejection rates of malw. and benign traces.}
\label{table:rejection_tpr}
\vspace{-0.08in}
\end{table}

\subsection{Evaluating the \sepep Scenario}
For our models in \sepep, we select up to 4 EP traces per sample to form our training set.
As we will show in~\cref{sec:environment_shift}, the behavior variability among the real-world traces of a sample~\cite{avllazagaj2021malware} impacts the model's predictions.
Consequently, including more traces per sample for training provides better coverage of the behaviors in the wild.
We find that using more than 4 traces per sample provides diminishing returns, \eg for 1, 2, 4, and 16 traces per sample, the TPR of \maldy is 43.0\%, 44.8\%, 49.5\%, and 49.9\%, respectively.

In Table~\ref{table:tpr_results}, we see a much higher EP performance in \sepep than in \ssbep{}, as expected.
Although training and testing sets are from similar distributions in \sepep (as opposed to \ssbep{}), the performance still lags behind the \ssbsb{}{} performance, \eg 95\% vs. 50\% TPR and 99\% vs. 88\% AUC (Table~\ref{table:app_auc_results}).
In the following sections, we shed light on this discrepancy and what makes endpoint detection particularly more challenging than a sandbox-only scenario.

\takeaways{Distribution shift causes ML methods trained on sandbox traces to perform poorly on endpoint traces (70+\% TPR drop). Even when trained on endpoint traces, these methods' endpoint performance still cannot reach their sandbox-based performance ($\sim$20\% vs. 95\% TPR), hinting at the inherent challenges in endpoint detection.}

\section{Endpoint Detection Deals With Difficult-To-Classify Samples}
\label{sec:sample_shift}

As described in~\cref{ssec:detection_workflows}, an endpoint detector operates on samples that cannot be classified statically---a subset of all samples in the wild.
The filtering effect this introduces on the distribution of samples relevant to endpoint detection has not been systematically measured before.
To illustrate, we examine (based on Table~\ref{table:top_fams_pubs}) the top malware families in our EP test set (undetected samples) and the SB test set (the samples from public corpora).
Although both datasets include only the samples first seen in 2017-2018, they encode different priors over samples.
For example, generic malware (\ie malware that could not be placed into any known family) covers 23.8\% of the EP samples vs. 4.8\% of the SB samples.
In the EP set, there are families such as \texttt{Khalesi} and \texttt{Emotet} that were rampant in 2018~\cite{emotet}.
Conversely, in the SB set, we see older families such as \texttt{Sivis} and \texttt{Upatre} (circa 2013-2014), whose variants still spread to this day.
Moreover, in Table~\ref{table:top_fams_pubs}, we share top benign sample publishers (extracted from the code-signing certificates when available), indicating a similar distribution difference in benign samples.
Overall, the typical distribution of training and testing samples used by prior work in \ssbsb{}{}~\cite{karbab2019maldy,jindal2019neurlux,dambra2023decoding} is substantially different than the distribution of testing samples in \ssbep{} and \sepep.

\topic{Borderline Samples and Label Noise}
A common practice in related work in malware detection with ML is discarding \emph{borderline} samples on which the anti-malware engines on VirusTotal are not in consensus.
Different works use different criteria; for example, they discard a sample if it is detected by more than zero and less than five engines~\cite{kuchler2021does}, 20 engines~\cite{mantovani2020prevalence} or 40 engines~\cite{lucasadversarial}.
Even works that do not explicitly discard borderline samples may follow a biased data collection process.
For example, recent related work has suggested that dynamic features have limited benefits over static features~\cite{dambra2023decoding}.
The benign samples used in this work were from a trusted Windows software repository, and 93\% of their malware samples were detected by 20+ engines.

We consider the samples detected by more than zero and less than 20 engines borderline.
This results in 27.7\% borderline samples in our EP test set (vs. 15.9\% in the SB test set).
The prior practice would have discarded over a quarter of the samples that a real-world endpoint detector encountered.
Next, we will show that this artificially inflates the measured success of a model.
As described in~\cref{ssec:datasets}, we do not discard any samples and label a sample as malware if more than five engines detect it.
This threshold poses a trade-off for the ground truth: as it increases, more malware samples may be labeled as benign, and as it decreases, more benign samples may be labeled as malware.
We present a case study in~\cref{ssec:dist_case_studies} to demonstrate how this trade-off plays out.

\subsection{The Impact of Sample Distributions}

Here, we measure how the distribution of samples in the test set changes a model's measured (and perceived) performance.
To this end, we resample our test sets according to some criteria, \eg following the malware family distribution in the EP test set, and evaluate our models on these new test sets.
This simulates different distributions over a model's test samples and allows us to do controlled experiments.
We focus on \maldy models due to their superior EP performance in~\cref{sec:sb_limitations}.

\begin{table}[h]
\centering
\normalsize
\begin{adjustbox}{max width=0.98\columnwidth}
\begin{tabular}{lp{0.83cm}p{0.83cm}||p{0.83cm}p{0.83cm}||p{0.83cm}p{0.83cm}}
\toprule
\multirow{2}{*}{\resultno{\#}} & \multicolumn{2}{c||}{\ssbsb{1}{1}} & \multicolumn{2}{c||}{\ssbep{1}} & \multicolumn{2}{c}{\sepep} \\
& TPR & AUC & TPR & AUC & TPR & AUC  \\
\midrule
\small\resultno{0} & \multicolumn{6}{c}{ { \textbf{\small No Resampling - Orig. Distributions}} } \\
& 93.2 & 99.0 & 16.7 & 78.4 & 49.5 & 87.5 \\
\midrule
\small\resultno{1} & \multicolumn{6}{c}{ { \textbf{\small Orig. Distributions w/o Generic Malware }} } \\
& 95.6 & 99.5 & 18.6 & 81.4 & 56.8 & 91.4 \\
\midrule
\small\resultno{2}& \multicolumn{6}{c}{ { \textbf{\small Only Generic Malware}} } \\
& 38.4 & 85.8 & 9.8 & 67.8 & 23.9 & 74.2 \\
\midrule
\small\resultno{3}& \multicolumn{6}{c}{ { \textbf{\small Discard Borderline Samples}} } \\
& 96.7 & 99.5 & 22.6 & 83.9 & 66.9 & 94.2 \\
\midrule
\small\resultno{4}& \multicolumn{6}{c}{ { \textbf{\small Malw. Resampled Following EP Test (w/ Generic)}} } \\
& 63.2 & 94.2 & 16.7 & 78.4 & 49.5 & 87.5 \\
\midrule
\small\resultno{5}& \multicolumn{6}{c}{ { \textbf{\small Malw. Resampled Following EP Test (w/o Generic)}} } \\
& 72.1 & 97.2 & 18.6 & 81.4 & 56.8 & 91.4 \\
\bottomrule
\end{tabular}
\end{adjustbox}
\caption{The impact of sample distributions in the test set on model evaluations. \resultno{0-5} denote different experiments.}
\label{table:family_shift}
\vspace{-0.11in}
\end{table}

Table~\ref{table:family_shift} presents the results of our resampling experiments.
We resample the test sets according to five criteria---denoted as \resultno{0--5} where \resultno{0} is 
testing on the original test set of each respective scenario.
In \resultno{1}, leaving out generic malware samples from original test sets leads to a significant boost in performance, whereas keeping only generic malware in \resultno{2} causes a massive drop.
This aligns with prior claims that generic malware is at the borderline and harder to classify~\cite{ugarte2019close}.
The fact that \texttt{Generic} has $\sim5\times$ higher coverage in the EP test set than in the SB test set (23.8\% vs. 4.8\%) also demonstrates the filtering effect that funnels more difficult samples to the endpoint detector.
Next, in \resultno{3}, we discard the borderline samples (identified in the previous section) from all test sets, simulating a common prior practice~\cite{kuchler2021does, mantovani2020prevalence,dambra2023decoding}.
This yields a significant boost in all scenarios: 5.9\% TPR boost in \ssbep{1} and 17.4\% in \sepep.
In \resultno{4} and \resultno{5}, we resample the test sets to match the distribution of malware families in the EP test set, with and without generic malware, respectively.
Note that, for \ssbep{1} and \sepep, \resultno{4} is equivalent to \resultno{0} and \resultno{5} is equivalent to \resultno{1}.
We observe a large decrease in the original \ssbsb{1}{1} performance, \eg from 93.2\% TPR to 63.2\% in \resultno{4} and to 72.1\% in \resultno{5}.
Despite neither including any generic malware, the \ssbsb{1}{1} performance in \resultno{5} is significantly lower than \resultno{1}.
This suggests that non-generic malware samples relevant to endpoint detection are generally harder to classify correctly, whether based on their sandbox or endpoint traces.
Consequently, the performance of prior behavioral detectors may be overestimated as they are evaluated on samples that are not representative of the distribution of samples faced in endpoint detection.
Although the distribution of malware samples explains most of the performance loss when ML classifiers are applied to EP traces, an unexplained gap remains (63.2\% vs. 49.5\%), which we attempt to explain in~\cref{sec:environment_shift}.

\subsection{Case Studies}
\label{ssec:dist_case_studies}

Next, we present two case studies illustrating sample distribution challenges in endpoint malware detection.

\vspace{0.05in}

\topic{Difficult Benign Samples}
Table~\ref{table:top_fams_pubs} lists Microsoft as the top benign publisher in our EP test set, covering 7.8\% of all benign samples.
This is counterintuitive as Microsoft is a trusted publisher, and our EP dataset only records samples that could not be classified statically as malware or benign.
We discovered that almost all (98\%) Microsoft samples in our EP test set share the \amdelta prefix in their filenames, which correspond to periodic patches to Windows Defender.
The \maldy model in \sepep outputs an average score of 37\% on \amdelta endpoint traces, almost 2.5$\times$ of the average score across all benign traces.
Note that the score quantifies the model's confidence that the input trace belongs to a malware sample.
In \sepep, on a test set containing only \amdelta samples as the benign samples, the model achieves only 62.6\% AUC, compared to 87.5\% with all benign samples.
Moreover, there are reports~\cite{amdeltareport} about anti-malware software falsely flagging \amdelta files.
A popular configuration repository for Sysmon---a tool to monitor process activities---includes an allow-list rule specifically for \amdelta~\cite{amdelta_syzmon}.
These suggest that \amdelta samples are difficult to classify correctly as they make sensitive modifications on hosts.
In contrast, our SB test set contains only two \amdelta samples (less than 0.01\% of all benign samples).
This highlights that an endpoint detector often faces difficult---false positive prone---benign samples in addition to difficult malware families, which is a root cause behind poor performance in \ssbep{} and \sepep.

\topic{Label Noise in Borderline Samples}
\roblox is a popular game creation platform.
Our EP train and test sets contain 19 and 29 samples from \roblox.
Despite being from a trusted publisher, there are reports about anti-malware products flagging \roblox samples.
Six (31\%) of our 19 training samples were labeled as malware because they were detected by over five engines on VirusTotal, whereas none of the 29 test samples were labeled as malware.
This causes a model to associate \roblox samples with malware-ness during training, potentially introducing false positives during testing.
For example, our \maldy model in \sepep outputs an average score of 40\% on the endpoint test traces from \roblox samples.
This demonstrates how borderline samples such as \roblox can introduce label noise, particularly in endpoint detection, where they are more common.

\subsection{Combating Sample Distribution Challenges}
\label{ssec:resampling}

Building on our observations, we propose two strategies to improve the performance in \ssbep{} and \sepep.

\vspace{0.05in}

\topic{Soft-Labeling Against Label Noise in \sepep}
Malware detectors are generally trained using hard binary labels---0 for benign and 1 for malware.
However, hard labels are hazardous when noisy as they force the model to overfit to a wrong prediction~\cite{li2019learning}.
The high frequency of borderline, \ie potentially noisy, samples exacerbates this problem in endpoint detection.
Research suggests that \emph{soft labels} can be effective against label noise~\cite{lukasik2020does} by preventing the model from getting too confident on noisy labels during training.  
To improve the \sepep performance, we implement a function to assign soft labels to our borderline training samples.

For a given sample $P_i$, our function computes its soft label as ${y_i\seq\text{min}\left( d_i^\theta/\beta^\theta, 1 \right)}$, where $d_i$ is the number of VirusTotal engines that detected $P_i$.
This function outputs $y_i\seq0$ when $d_i\seq0$ (confident benign), and it saturates at $y_i\seq1$ when $d_i{\geq}\beta$ (confident malware).
The hyper-parameter $\theta$ determines how fast $y_i$ grows from $0$ to $1$ as $d_i$ increases.
We present three curves this function generates when $\beta\seq20$ in Figure~\ref{fig:soft_label_func}.
We set $\beta\seq20$ and $\theta\seq0.75$ without careful tuning and train new models on this soft-labeled EP training set.
The results in Table~\ref{table:resampling_softlabeling} (\emph{first} segment) show a significant gain for \neurlux (5.7\% higher TPR) and a moderate gain for \maldy (1.1\% higher TPR).
This suggests that \neurlux is more vulnerable to label noise, which is expected as \neurlux is a more complex architecture than \maldy with a higher capacity for overfitting.
These improvements support our intuition regarding label noise and its effect on ML-based detectors.
We leave the exploration of advanced methods against label noise, such as semi-supervised learning~\cite{wu2023grim}, to future work.

\begin{table}[hbt]
\centering
\normalsize
\begin{adjustbox}{max width=0.98\columnwidth}
\begin{tabular}{p{0.75cm}p{0.75cm}p{0.75cm}p{0.75cm}||p{0.75cm}p{0.75cm}p{0.75cm}p{0.75cm}}
\toprule
\multicolumn{4}{c||}{{\textbf{\large Before}}} & \multicolumn{4}{c}{{\textbf{\large After}}} \\
\multicolumn{2}{c}{\maldy} & \multicolumn{2}{c||}{\neurlux} & \multicolumn{2}{c}{\maldy} & \multicolumn{2}{c}{\neurlux} \\
TPR & AUC & TPR & AUC & TPR & AUC & TPR & AUC \\ 
\midrule
\multicolumn{8}{c}{\textbf{Soft Labeling} (\sepep) }  \\
49.5 & 87.5 & 43.5 & 86.8 & 50.6 & 87.8 & 49.2 & 88.2 \\ 
\midrule
\multicolumn{8}{c}{\textbf{Resampled Training Set --- Uniform} (\ssbep{1}) }  \\
16.7 & 78.4 & 10.2 & 74.1 & 15.9 & 77.2 & 8.6 & 74.0 \\ 
\midrule
\multicolumn{8}{c}{\textbf{Resampled Training Set --- EP Training} (\ssbep{1}) }  \\
16.7 & 78.4 & 10.2 & 74.1 & 20.0 & 78.6 & 11.4 & 74.9 \\\bottomrule
\end{tabular}
\end{adjustbox}
\caption{Our strategies to improve the performance against sample distribution challenges in endpoint detection.}
\label{table:resampling_softlabeling}
\vspace{-0.04in}
\end{table}

\topic{More Accurate Training Distributions in \ssbep{}}
Our models in \ssbep{} were trained on the distribution of samples captured in public corpora.
Although this is standard for obtaining the best \ssbsb{}{} performance~\cite{jindal2019neurlux,trizna2023nebula}, it is not ideal for \ssbep{} where the test set follows a different distribution.
ML models are heavily influenced by the priors encoded in their training sets because the empirical risk minimization (ERM) paradigm minimizes the average loss over all training samples. 
For example, 10\% of the malware in the SB training set is from the \texttt{Virut} family, which will be prioritized over \texttt{Emotet}, which accounts for 0.3\%.

To alleviate this discrepancy in \ssbep{}, we turn our attention to ML research, where \emph{group robustness} techniques~\cite{sagawa2020investigation,liu2021just,zhang2022contrastive} have been effective against a similar problem.
These techniques aim to train models that perform well on each subgroup (\eg demographic groups) instead of favoring groups over-represented in the training set.
We implement a prior group robustness method~\cite{cao2019learning} by creating an SB training set with an equal number of samples from each malware family.
Unfortunately, as presented in Table~\ref{table:resampling_softlabeling} (\emph{second} segment), this uniform sampling strategy ends up hurting the performance in \ssbep{}.
We attribute this shortcoming to the large number of groups (families) in our problem, compared to standardized ML benchmarks, over 600 vs. typically less than 10~\cite{liu2021just,kirichenko2023last}. 
Moreover, malware family labels are often incomplete or noisy~\cite{sebastian2020avclass2,kotzias2019mind}.
For example, in our SB1 training set, there are many noisy malware family tags such as \texttt{vbkryjetor} (13
samples). 
As a result, the model is forced to balance hundreds of groups, which also might be inaccurately tagged.
These results highlight a key difference between real-world security applications and simplified ML benchmarks, on which most solutions are developed and evaluated, raising doubts about whether an \emph{uninformed} training strategy could work.

Finally, we experiment with an \emph{informed} strategy where the \ssbep{} training set is resampled to follow the family distribution in the EP training set.
Widely used threat intelligence platforms can guide practitioners about which families to include in training~\cite{ciscothreatintel}.
Table~\ref{table:resampling_softlabeling} (\emph{third} segment) shows that this strategy brings moderate improvements: 3.3\% and 1.2\% higher TPR for \maldy and \neurlux, respectively.
We will use this strategy to train models in the rest of our paper.
%

\takeaways{The distribution of samples selected evaluate malware detectors must account for the detector's position in a pipeline. Accordingly, creating realistic test sets for ML-based behavioral detectors (\ie the final step in the pipeline) reveals the overestimated performance in prior evaluations done on broad distributions (63\% vs. 93\% TPR).}

\section{Environment-Sensitive Program Behaviors Hurt Endpoint Detection}
\label{sec:environment_shift}

We have shown the performance implications of the distribution of samples encountered by endpoint detectors.
However, even when we accounted for distribution differences (\resultno{4} in Table~\ref{table:family_shift}), the performance in \ssbsb{}{} (63.2\% TPR) is higher than the performance in \ssbep{} (16.7\% TPR) and in \sepep (49.5\% TPR).
In this section, we aim to illuminate the remaining performance gap by studying the impact of variable program behaviors on ML-based approaches.

It is known that program behaviors are environment sensitive and, as a result, a sample can exhibit varying behaviors in different environments~\cite{balzarotti2010efficient,lindorfer2011detecting,avllazagaj2021malware}.
Malware samples are more environment-sensitive than benign samples because, for example, they execute only if the host has specific software vulnerabilities~\cite{avllazagaj2021malware} or do not execute if they detect they are running in a sandbox~\cite{lindorfer2011detecting,miramirkhani2017spotless}.
Building on these insights, we aim to characterize the impact of variable program behaviors on ML-based detectors.
We focus on the variability stemming from environment differences between sandboxes and endpoint hosts (for the \ssbep{} scenario) and between different endpoint hosts (for the \sepep scenario).
We perform our experiments using the improved \maldy models from~\cref{ssec:resampling}.

\subsection{The Diversity Among Endpoint Traces}

Table~\ref{table:distances} shows how much traces vary when a set of related samples (\eg from the same malware family) is run in different environments.
We aggregate all the traces from three malware families and two benign publishers in our SB1, SB2, and EP datasets.
To cover a range of samples, we select \texttt{Wannacry} (ransomware), \texttt{Emotet} (banking trojan), and \texttt{Khalesi} (info-stealer) as the families, and \texttt{Opera} (web browser), \texttt{Roblox} (game platform) as the publishers.
We report the average pairwise distances between these traces within an environment, \eg between the traces from SB1 (\emph{SB1-SB1}), and between two environments, \eg between the traces from SB1 and EP (\emph{SB1-EP}).
We use normalized compression distance, which has been used in similar contexts~\cite{bailey2007automated,avllazagaj2021malware}.

\begin{table}[hbt]
\centering
\begin{adjustbox}{max width=0.99\columnwidth}
\large
\begin{tabular}{l|rrr||rrr}
\toprule
& \multicolumn{3}{c||}{\textbf{\large Within Envs.}} & \multicolumn{3}{c}{\textbf{\large Between Envs.}} \\
& \multicolumn{1}{c}{\normalsize {SB1-SB1}} & \multicolumn{1}{c}{\normalsize {SB2-SB2}} & \multicolumn{1}{c||}{\normalsize {EP-EP}} &\multicolumn{1}{c}{\normalsize {SB1-SB2}} & \multicolumn{1}{c}{\normalsize {SB1-EP}} & \multicolumn{1}{c}{\normalsize {SB2-EP}} \\
\midrule
Wcry & 0.16 & 0.11 & 0.44 & 0.28 & 0.67 & 0.50 \\
Emot & 0.48 & 0.35 & 0.50 & 0.55 & 0.65 & 0.53 \\
Khal & 0.33 & 0.29 & 0.48 & 0.52 & 0.62 & 0.57 \\
\midrule
Oper & 0.05 & 0.07 & 0.69 & 0.43 & 0.70 & 0.75 \\
Rblx & 0.18 & 0.04 & 0.63 & 0.38 & 0.67 & 0.69 \\
\bottomrule
\end{tabular}
\end{adjustbox}
\caption{Avg. pairwise distances between traces of samples from three malware families and two benign publishers.}
\label{table:distances}
\vspace{-0.06in}
\end{table}

We observe: \bnumber{i} traces from the same sandbox are almost always very similar; \bnumber{ii} traces from two endpoint hosts are dissimilar (also shown in~\cite{avllazagaj2021malware}); \bnumber{iii} traces from different sandboxes are dissimilar but much less so than one trace from a sandbox and one trace from an EP host.
The observation \bnumber{iii} can be detrimental for \ssbep{}, as the features learned from SB traces might be irrelevant to EP traces, and \bnumber{ii} is potentially detrimental for both scenarios.
The dissimilarity between SB and EP traces implies that a model's training set in \ssbep{} follows a different distribution than its test set, explaining the high drift detection in~\cref{ssec:eval_sbep}.

\begin{figure}[hbt]
    \centering
    \includegraphics[width=0.9\linewidth]{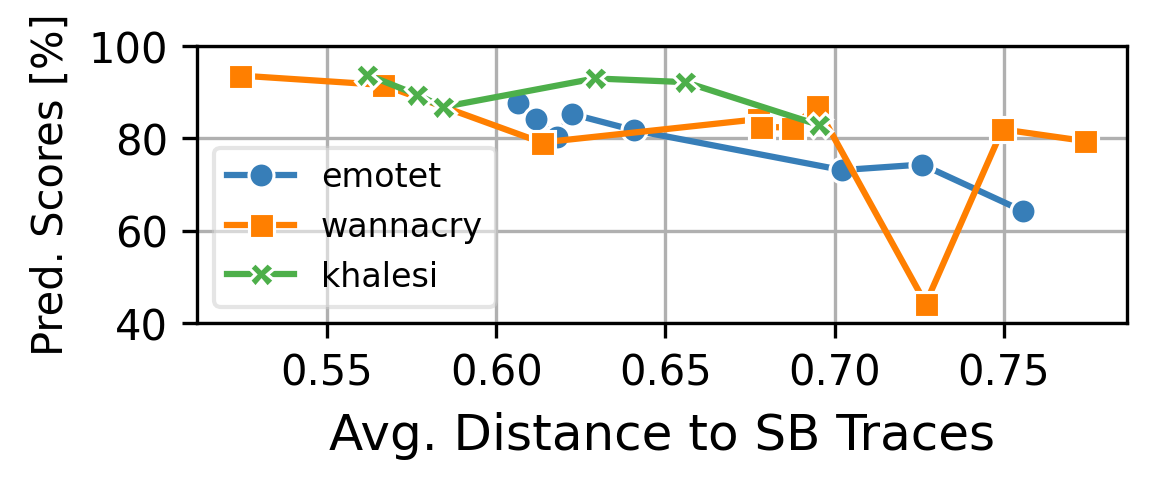}
    \caption{Relationship between the avg. distance of an input EP trace to the SB traces and the model's output score in \ssbep{}. Each marker represents the avg. of 30 input traces.}
    \label{fig:fam_distance_preds}
  \vspace{-0.15cm}
\end{figure}

Next, we demonstrate how \bnumber{iii}---the distance between SB traces and EP traces---affects a model in \ssbep{1}.
Figure~\ref{fig:fam_distance_preds} shows that the model is more confident for traces that are more similar to SB traces.
Here, each marker represents a group of 30 traces of a malware family from endpoint hosts, and the x-axis quantifies the average distance of these traces to all traces of this family from SB1.
The lower this distance, the more similar a group of endpoint traces is to traces of this family from SB1.
For example, the model outputs a 90\% score for \texttt{Emotet} EP traces with an average distance of 0.60 to \texttt{Emotet} SB1 traces, which drops to 60\% for an EP trace with a distance of 0.75.
Consequently, the less similar the input EP trace is to the SB traces in the training set, the worse decisions the model makes and performs poorly in \ssbep{}.

In line with our observation \bnumber{ii}, research shows that a sample's behaviors vary across hosts due to host-related or external factors~\cite{avllazagaj2021malware}.
To study how this variability impacts ML, we define $S_{i,h}$ as the set of a model's prediction scores on the EP traces of a sample $P_i$ collected within $h$ hours after $P_i$ was first seen---${S_{i,h} = \{f(x_{i,j}) \; \lvert \; x_{i,j} \in \mathbf{x}^0_{i,\:(t<h)} \}}$.

\begin{figure}[hbt]
    \centering
  \begin{subfigure}[b]{0.49\columnwidth}
    \includegraphics[width=\linewidth]{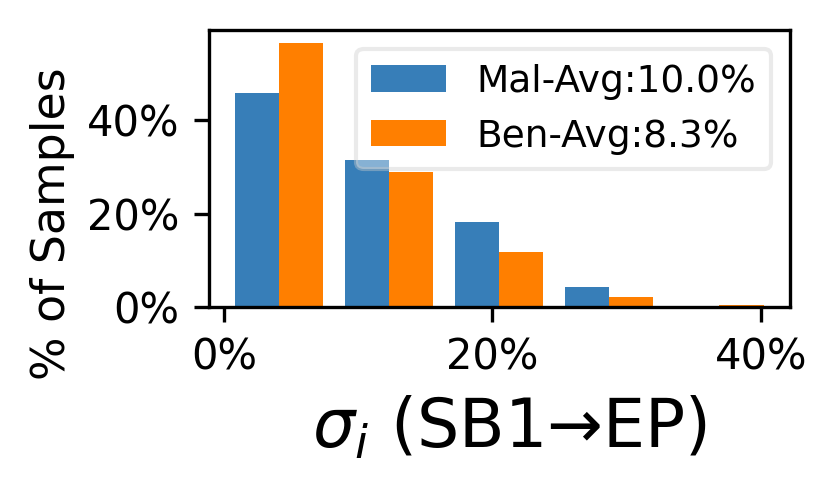}
  \end{subfigure}
  \begin{subfigure}[b]{0.49\columnwidth}
    \includegraphics[width=\linewidth]{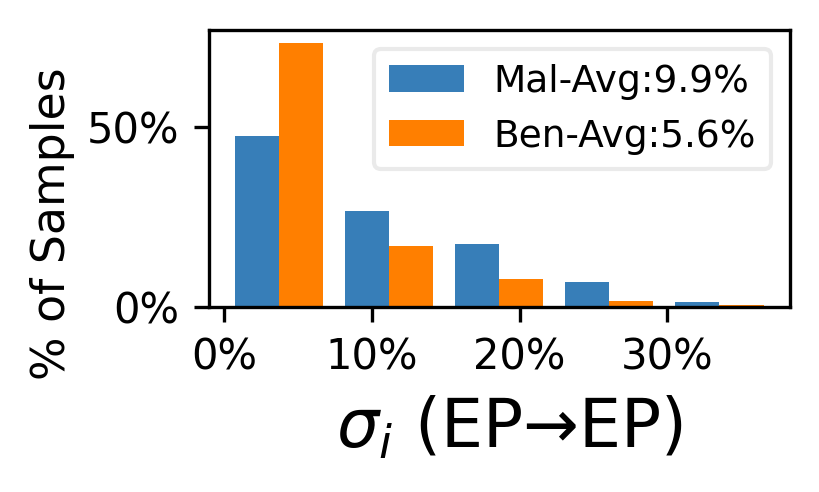}
  \end{subfigure}
\caption{Prediction score std. deviations on the EP traces of each sample in the EP test set; in \ssbep{1} and \sepep.} 
    \label{fig:ep_std_devs}
  \vspace{-0.07in}
\end{figure}

Figure~\ref{fig:ep_std_devs} presents two histograms of the standard deviations of $S_{i,\:h\seq24}$ (denoted as $\sigma_i$), measured over the samples in the EP test set.
With $\sigma_i$, we quantify the robustness of a model's predictions to the variability among different EP traces of $P_i$.
First, score variability is higher in \ssbep{}, indicating that a model trained on sandbox traces is more sensitive to behavior variability in the wild.
Further, the average $\sigma_i$ for malware is higher than benign (9.9\% vs 5.6\% in \sepep), aligning with the finding that malware samples behave more variably across hosts~\cite{avllazagaj2021malware}.
Score variability is also not uniform across samples; some have nearly zero, and some have over twice the average.
This is dictated by the malware type, \eg botnets execute custom commands, resulting in more variable behaviors~\cite{avllazagaj2021malware} and higher score variability.
Overall, behavior variability can manifest as score variability and trigger errors (\eg some traces of a malware sample classified as benign), and ML-based detectors are not naturally robust to it.

\topic{\modification{The Impact of Sandbox Configurations}}
Our experiments use sandbox traces obtained from commercial sandboxes with configurations unknown to us, likely optimized to identify malware from its sandbox behaviors.
The configuration differences between sandboxes and endpoint hosts (\eg operating systems, software versions) lead to the large dissimilarity between SB and EP traces we observed.
This raises the question of whether a sandbox can be configured to obtain traces that approximate those from real-world hosts.
Training on such sandbox traces will likely result in higher performance in \ssbep{}.
We acknowledge this as a valid possibility, but it remains an open research question in the current landscape.
There is no accepted standard to configure sandboxes in such a manner, and some research suggests that it would be challenging to replicate endpoint systems, \eg simulating user interactions~\cite{liu2022enhancing} or diverse real-world configurations~\cite{mills2020investigating}.
%
Nonetheless, currently, the research community collects training sets mainly from a single sandbox for ML-based detection solutions~\cite{miller2016reviewer,karbab2019maldy,trizna2022quo}.
We are unable to collect our own sandbox traces with various configurations to analyze how sandbox configurations affect detection performance, as this would cause a disparity between our endpoint and sandbox data (explained in~\cref{ssec:datasets}).
We highlight this as an important direction for future work.

\subsection{Case Studies}
Next, we present two case studies to illustrate the nature of behavior variability and its impact on ML models.

\topic{Benign Behavior Variability}
\texttt{RobloxLauncher} and \texttt{OperaPatcher} are two benign samples on the opposite ends of the behavior variability spectrum.
\texttt{RobloxLauncher} accepts user input through a graphical interface, such as mouse clicks on menu items.
We clustered the EP traces of this sample to discover two main execution paths: \bnumber{i} downloading a long list of assets (such as \texttt{.mp3} or \texttt{.jpeg} files), presumably to update the game; \bnumber{ii} creating temporary Internet files and starting a process, presumably for launching the game.
Our \sepep model makes more accurate predictions on the traces in \bnumber{i} than \bnumber{ii} (26\% vs 38\% average prediction score).
In its SB2 trace, the sample downloads a similar list of assets, whereas, in its SB1 trace, it creates only a log file and stops, likely because it could not access the Internet.
Neither sandbox triggers the game launch execution path, so we expect a model trained on sandbox traces to struggle to classify endpoint traces that launch the game.
Applied to a model trained on SB1 traces, a drift detector (\cref{ssec:eval_sbep}) consistently rejects the EP traces of this sample.
We believe this illustrates a broader challenge: emulating user interactions in sandboxes is challenging~\cite{liu2022enhancing}, causing major differences in the data distribution of sandbox vs endpoint traces from benign samples.
Conversely, \texttt{OperaPatcher} performs almost the same actions in all its SB1, SB2, and EP traces: creating localization and library files to patch a browser.
This sample executing without any user interaction eliminates behavior (and score) variability across traces.

\topic{Non-Malicious Malware Traces}
In behavioral malware detection, most commonly, the label of a sample (\eg obtained from VirusTotal) is transferred to all traces from this sample, which are then used for training and testing.
This practice implicitly assumes that \emph{all} traces of a malware sample contain some discernible malicious activity.
However, this can be problematic in cases where the sample has refused or failed to execute, \eg because its remote infrastructure is down~\cite{yong2021inside}.

To assess this approach, we study the \texttt{Wannacry} ransomware family.
We select \texttt{Wannacry} as it is thoroughly dissected (unlike most families), and its indicators-of-compromise (IOCs) are well-known, allowing us to gauge whether a particular trace is associated with compromise.
Using multiple sources, we create a list of IOCs for \texttt{Wannacry} and look for them in our SB1, SB2, and EP traces.
We split the traces into two sets based on whether they contain any IOC and then measure the average prediction scores of our models on each set.
Based on Table~\ref{table:wannacry_iocs}, we observe \bnumber{i} most traces contain at least one IOC (\eg 97.7\% of the EP traces), \bnumber{ii} the model's average prediction score is lower on the traces with no IOC (\eg 97\% vs. 52\% in \sepep), which is still significantly higher than the averages on benign traces (13\%).
The ability to associate malware traces with no IOC with maliciousness is a strength of ML-based methods over IOC-based detectors.
Any malware that reaches the execution stage is hazardous, regardless of whether it successfully compromises the host.

\subsection{Spurious Features From Sandboxes}
\label{ssec:sandbox_artifacts}
We have shown that sandbox and endpoint traces for a given sample \emph{are} different.
Next, we examine \emph{why} they are different and \emph{how} this difference may trigger unreliable predictions from the ML models in the \ssbep{} scenario.

\topic{Sandbox-Specific Artifacts}
We found several sandbox-specific features that are highly prevalent (seen in many traces) and predictive (seen only in malware traces) but occur only in one sandbox and not in the other environments.
ML models typically exploit such features to minimize the loss and may overlook other predictive features~\cite{pezeshki2021gradient}, causing the model to perform poorly on traces collected from a different sandbox or from endpoint hosts.
Table~\ref{table:sandbox_artifacts} presents four of these features (called \emph{artifacts}) that we have found in SB1 and SB2 traces.
We have identified over 100 such artifacts, most frequently co-occurring in traces.
%
%
For each artifact, we report its prevalence (\textbf{Prv.}) and its malware ratio (\textbf{MalR.}) in SB1, SB2, and EP training sets.
Prv. is the percentage of samples in which the artifact is present, and MalR. is the percentage of these samples labeled as malware.

\begin{table}[hbt]
\large
\centering
\begin{adjustbox}{max width=0.90\columnwidth}
\begin{tabular}{lrr|rr|rr}
\toprule
 \multirow{2}{*}{\Large File Name} 
 & \multicolumn{2}{c|}{\textbf{SB1 Traces}} & \multicolumn{2}{c|}{\textbf{SB2 Traces}} & \multicolumn{2}{c}{\textbf{EP Traces}} \\
& Prv. & MalR. & Prv. & MalR. & Prv. & MalR. \\
\midrule
 \texttt{SogouExp.} & 9.0\% & 100.0\% & 0.0\% & --- & 0.2\% & 0.0\% \\
\texttt{PersonalB.} & 3.3\% & 100.0\% & 0.0\% & --- & 0.0\% & --- \\
\midrule
\texttt{Spotify} & 0.1\% & 0.0\% & 3.0\% & 100\% & 0.2\% & 0.0\% \\
\texttt{Python} & 0.0\% & --- & 3.1\% & 100\% & 0.1\% & 0.0\% \\
\bottomrule
\end{tabular}
\end{adjustbox}
\caption{Some strong malware features found in specific sandboxes that do not generalize to other environments.}
\label{table:sandbox_artifacts}
\vspace{-0.04in}
\end{table}

These artifacts appear to be a result of the specific configuration and particular apps pre-installed on the specific sandbox (see~\cref{app:artifacts_dive}).
As real-world hosts and other sandboxes do not share the same configuration, this explains why an endpoint classifier trained on traces from one sandbox might fail to generalize to endpoint traces seen in the wild.

\topic{The Impact of Sandbox Evasion}
Sandbox evasion is a common mechanism malware authors use to make a sample's sandbox trace dissimilar to its real-world traces.
If running in a sandbox, evasive malware attempts to avoid analysis or terminate early~\cite{mcafeeevasion,miramirkhani2017spotless}.
In~\cref{app_sandbox_evasion}, we present a series of quantitative experiments to understand the implications of sandbox evasion for ML-based detectors.
We find that an ML-based detector trained on sandbox traces predicts a higher score of maliciousness for the shortest traces.
This \emph{length bias} boosts the performance in the \ssbsb{}{} scenario, where very short traces are more likely from malware.
The same bias, however, hurts the performance in the \ssbep{} scenario, where very short traces are not more likely from malware.
This makes trace length a spurious correlation learned from sandbox traces that fails to generalize to endpoint traces.
Our evidence suggests that this correlation is introduced by evasive malware, which tends to produce short sandbox traces~\cite{kuchler2021does}.

\subsection{Invariant Learning to Counter Environment-Sensitivity}
\label{ssec:env_invariance}
We have shown that the environment-sensitivity of program behaviors impacts the model.
When the classifier exploits features that are specific to the training environment (\eg a sandbox artifact), its predictions fail to generalize.
Table~\ref{table:sandbox_artifacts} suggests that the artifacts of SB1 are absent in SB2, and vice versa, as they are configured sufficiently differently.
This highlights an opportunity: learning features that are invariant across two environments (\eg both sandboxes) might yield better generalization to endpoint detection.

Prior work has studied invariant learning in self-supervised learning, \emph{e.g.}, viewpoint-invariant visual features~\cite{purushwalkam2020demystifying}.
We employ the Siamese loss~\cite{chen2021exploring}, which forces the model to produce similar embeddings (measured by a distance metric) on pairs of related inputs.
In our case, each pair consists of two traces of the same sample from different sandboxes (in \ssbep{}) or from different EP hosts (in \sepep).
We aim to make the model invariant to the differences in these pairs (such as sandbox-specific artifacts) to gain robustness to variable behaviors.
The critical question is whether invariance across SB traces translates to invariance across EP traces.
We represent a pair of traces of a sample $P_i$ as ($x_{i,0}, x_{i,1})$ and the set of all pairs as $\mathbf{D}$; and define the following loss function that is added to the standard ERM loss as a regularizer:

\vspace{-0.1in}

\begin{gather*}
    \mathcal{L}_\text{inv} = \dfrac{1}{\lvert \mathbf{D} \rvert} \sum_{(x_{i,0}, x_{i,1}) \in \mathbf{D}} - \mathrm{cos} (\text{enc}(x_{i,0}), \text{enc}(x_{i,1})) \\
    \addlinespace
    \mathcal{L}_\text{final} = \mathcal{L}_\text{ERM} + \alpha \mathcal{L}_\text{inv}
\end{gather*}

Here, $\mathrm{cos}$ is the cosine similarity, and $\alpha$ controls the intensity of invariance regularization: if set too high, the embeddings might collapse into a single point.
In \ssbep{}, $\mathbf{D}$ contains the training traces in SB1$\cap$SB2 (Table~\ref{table:dataset_stats}), yielding $\sim$15K pairs.
In \sepep, we create $\mathbf{D}$ by randomly selecting pairs of traces from each sample in our EP training set ($\sim$88\% of the samples have more than one trace).
\modification{Note that our proposed solution for \ssbep{} relies on running a large set of samples in multiple sandboxes.
Although this is standard practice in malware analysis~\cite{yong2021inside}, it is typically not done for training ML models for behavioral detection.}

\begin{table}[hbt]
\large
\centering
\begin{adjustbox}{max width=0.95\columnwidth}
\begin{tabular}{lrrrrrr}
\toprule
Scenario & \multicolumn{1}{c}{$\alpha\seq0$} & \multicolumn{1}{c}{$\alpha\seq0.02$} & \multicolumn{1}{c}{$\alpha\seq0.08$} & \multicolumn{1}{c}{$\alpha\seq0.32$} & \multicolumn{1}{c}{$\alpha\seq1.28$} \\
\ssbep{1} & 20.0\% & 19.1\% & 20.5\% & \color{red}{\textbf{21.6\%}} & 20.6\% \\
\sepep  & 50.6\% & 50.6\% & 50.6\% & 51.1\% & \color{red}{\textbf{51.8\%}} \\
\bottomrule
\end{tabular}
\end{adjustbox}
\caption{The impact of $\mathcal{L}_\text{inv}$ on endpoint detection TPR.}
\label{table:invariant_benefits}
\vspace{-0.1cm}
\end{table}

We present the results in Table~\ref{table:invariant_benefits}.
Here, we start from ($\alpha\seq0)$ the improved \maldy models in~\cref{ssec:resampling}.
Both \ssbsb{}{} and \ssbep{} models mildly benefit (1--2\% TPR boost) from $\mathcal{L}_\text{inv}$, though at different values of $\alpha$.
We hypothesize that some environment-dependent features are still useful, \emph{e.g.}, when a malware family evades one sandbox but not the other, making excessive invariance undesirable.
We leave improvements, \emph{e.g.}, selective invariance to preserve useful features, or over more than two sandboxes, to future work.

\begin{figure}[hbt]
\centering
  \begin{subfigure}[b]{0.31\columnwidth}
    \includegraphics[width=\linewidth]{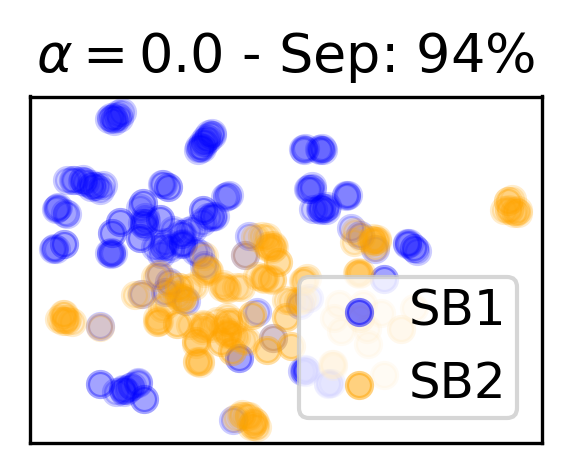}
  \end{subfigure}
  \begin{subfigure}[b]{0.32\columnwidth}
    \includegraphics[width=\linewidth]{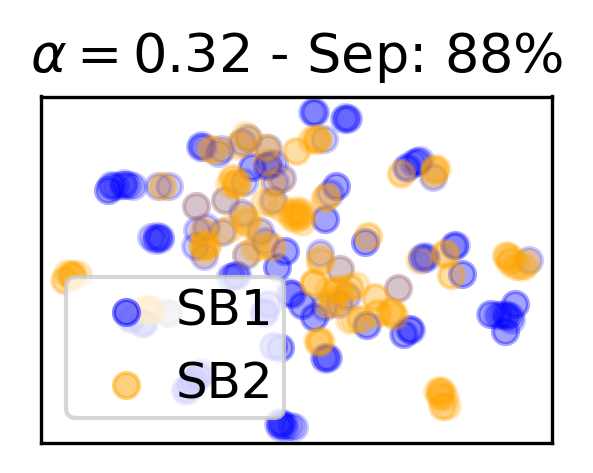}
  \end{subfigure}
  \begin{subfigure}[b]{0.32\columnwidth}
    \includegraphics[width=\linewidth]{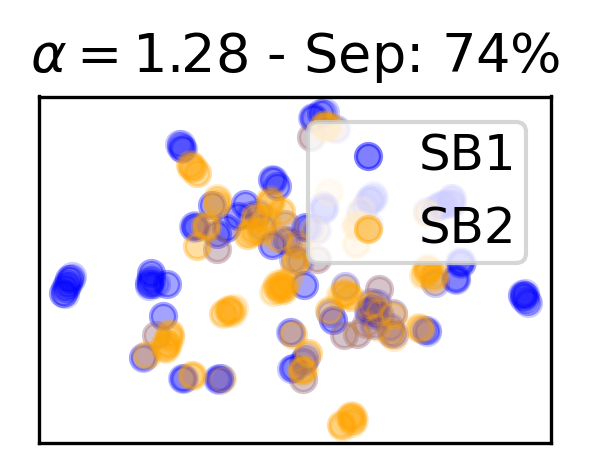}
  \end{subfigure}
\caption{The model's embeddings with increasing values of $\alpha$ for invariant learning. \emph{Sep} quantifies the separability of the SB1 and SB2 embeddings. t-SNE~\cite{van2008visualizing} visualization.}
    \label{fig:invloss_impact}
  \vspace{-0.1cm}
\end{figure}

To ensure $\mathcal{L}_\text{inv}$ is working as intended, in Figure~\ref{fig:invloss_impact}, we compare the embeddings that \ssbep{1}models produce on the testing traces in SB1$\cap$SB2.
Increasing $\alpha$ brings the embedding distributions on SB1 and SB2 traces visually closer.
Quantitatively, this decreases the accuracy of an SVM classifier in separating these embeddings as coming from SB1 or SB2 (from 94\% to 74\%).
Moreover, Figure~\ref{fig:ep_std_devs_w_env} presents two histograms of score standard deviations for our invariant models in \ssbep{} and \sepep.
In both scenarios, the predictions on EP traces have become less variable (compared to Figure~\ref{fig:ep_std_devs}).
The average standard deviation in \sepep for benign samples decreases significantly (from 5.6\% to 0.8\%), whereas it remains the same for malware samples.
The invariant model has become robust to the variability across the traces of a benign sample, which might not be possible for malware samples whose behaviors vary more.
Although we compute $\mathcal{L}_\text{inv}$ using only SB1 and SB2 traces in \ssbep{}, the resulting environment-invariance has transferred to endpoint traces, giving us a performance boost.
The evidence shows that learning environment-invariant features is promising and can offset the negative impact of behavior variability on ML-based detectors.

\takeaways{An execution environment's specific configuration might introduce artifacts into program traces, hurting behavioral detectors' generalizability to other environments. Learning invariant features over a sample's traces from multiple environments (\eg two sandboxes) can enhance ML models' robustness to such artifacts, improving their performance in real-world hosts with diverse configurations.}
\section{Discussion and Limitations}
\label{sec:discussion}

\topic{\modification{Dataset Bias, Size and Label Noise}}
Our study has some limitations due to our endpoint data.
First, this data was collected from hosts that use a single vendor's anti-malware product.
Although we cannot rule out selection bias, the fact that these hosts are located in over 100 countries in both enterprise and consumer settings suggests that our results have broad applicability. 
Second, our data was collected six years ago.
This is a common limitation in malware studies as collecting large-scale, up-to-date real-world data is infeasible for researchers~\cite{yang2021cade,barbero2022transcending,chen2023continuous}.
Nevertheless, to our knowledge, our endpoint dataset (originally collected in~\cite{avllazagaj2021malware}) is the only dataset of endpoint malware behavior analyzed in the literature. 
We focus on the gaps between sandbox-based and endpoint detection and not on the specifics of the threats of the time, making our observations still relevant today.  
Third, our data consists of only Windows hosts and lacks network-related actions.
As 95\% of malware is aimed at Windows~\cite{windowsmalware} and ML-based detectors can perform well without network actions~\cite{jindal2019neurlux}, we do not expect this to affect our findings.

Moreover, our endpoint dataset is much smaller than our sandbox dataset and smaller than what is available to a security vendor, potentially causing us to underestimate the \sepep performance in practice.
However, consider Figure~\ref{fig:fine_tuning_sb} where we train models in \sepep using increasing proportions of our endpoint dataset.
Here, the performance resulting from using 25\%, 50\%, and 100\% of the data for training is 35\%, 41\%, and 46\% TPR, respectively.
This hints that even with larger endpoint dataset sizes, the performance in \sepep is unlikely to catch \ssbsb{}{} due to fundamental challenges we exposed (difficult-to-classify samples and behavior variability).

Finally, for a given sample, labels obtained from VirusTotal (VT) are initially noisy and stabilize after around a year~\cite{zhu2020measuring}.
In our experiments (see~\cref{ssec:datasets}), we use the oldest label available (\eg from old VT reports) for the training samples and the labels from VT reports collected four years later (in 2022) for the testing samples.
This methodology approximates the realistic conditions for deploying a malware model.
To gauge the impact of label noise on our measurements, we compare the labels of training samples from recent VT reports to the labels we used to train our models.
In the SB and EP training sets, old and new labels agree on 98\% and 97\% of the samples, suggesting that the effect of label noise is minimal.

\topic{Early Detection of Malware}
We study detectors that use the whole execution trace of a sample, and thus can only detect malware once it terminates.
Although vendors offer ways to undo the damage from malware after its execution, such as quarantining~\cite{quarantine}, it is more desirable to catch malware in its tracks, \emph{e.g.}, before a ransomware sample starts encrypting personal files.
In that task, a detector has strictly less information available to classify a sample, which makes our results an upper bound on the performance of early malware detection.

\topic{Fine-Tuning a Sandbox-Based Model on Endpoint Traces}
In~\cref{app:fine_tuning}, we present an experiment where we first train a \maldy model only on sandbox traces; we then fine-tune this model on increasing amounts of data from our EP training set.
This hybrid approach simulates a realistic scenario where low-cost sandbox data is supplemented with high-cost endpoint traces from the wild.
We implement two strategies: fine-tuning all layers of the model and fine-tuning only the last layer.
We find that adding a few EP traces improves performance, and when very few EP traces are available, fine-tuning outperforms training from scratch (3--5\% higher TPR).

\topic{Attacks Against ML}
All ML-based detectors, especially deep-learning-based ones, are subject to adversarial attacks due to their sensitivity to input perturbations~\cite{szegedy2014intriguing,grosse2017adversarial}.
We expect our models will also be vulnerable to such attacks, \emph{e.g.}, an adversary can inject dummy actions into the behaviors of their malware sample designed to fool a model.
We consider a defense against these attacks to be out of scope for this work; 
instead, we aim to show that naturally occurring pressures, such as environment variability, also greatly degrade the performance of ML-based malware detectors in the wild, posing a critical security and trustworthiness challenge in practice.

\topic{Promoting Future Research}
Our investigation relies on a dataset of endpoint traces from the wild provided by Avllazagaj et al., who collected and analyzed it in their prior work~\cite{avllazagaj2021malware}.
Using this data, we have pinpointed the salient challenges in ML-based endpoint malware detection.
Performing evaluations only on lab-based data can obscure these challenges and produce biased or unreliable solutions inapplicable to the real world~\cite{cherubin2022online,jacobs2022ai}. 
Although its terms of use prevent us from sharing Avllazagaj et al.'s dataset, we built a pipeline that allows the community to perform realistic evaluations of ML-based behavioral malware detectors, aiming to minimize this bias.
Anyone may submit to us a pre-trained detector and a feature extractor that converts a trace in the standardized format (\cref{ssec:datasets}) into an input to the detector.
Upon receiving a submission, we will evaluate it on our endpoint traces dataset and return the submitter a detailed performance report to guide their research.
Further, if the submitter wishes, we will publicly list their results and submission details on the leaderboard on our website (\href{https://malwaredetectioninthewild.github.io}{https://malwaredetectioninthewild.github.io}).
To assist participants, we will release the sandbox dataset we collected and the metadata for over 200K samples used in our work.
The details of these artifacts can be found in~\cref{app:leaderboard}.
Please refer to our website for technical details, data access requests, and detector submission instructions.
This pipeline offers a blueprint for security vendors to drive research without publicly sharing their sensitive data.
We hope to spearhead it into more applications of ML for security where evaluation problems frequently hinder real progress.

\topic{\modification{Implications for Future Work}}
As described in~\cref{ssec:detection_workflows}, malware detection in practice involves a pipeline of techniques.
We believe ignoring this pipeline and treating each component as a standalone solution causes critical evaluation problems.
Consider ML-based static malware detectors, an active research area~\cite{anderson2018ember,raff2018malware,dambra2023decoding}.
They are ideally evaluated after filtering out the test samples that can already be detected by the pipeline components preceding static ML, such as existing blocklists, allowlists, and static signatures.
However, to our knowledge, this is not an established practice, which might cause an overestimation of real-world performance (similar to our findings in~\cref{sec:sample_shift}).
Our endpoint data allowed us to isolate and realistically evaluate ML-based behavioral detectors, revealing previously unknown challenges.
This, however, still falls short of an ideal end-to-end evaluation.
Instead of measuring the success of individual components, quantifying the overall effectiveness of a pipeline yields more relevant insights for users.
For example, if static components detect 95\% of the active malware at a given time and the dynamic components detect 80\% of the remaining malware, the overall pipeline would be at 99\% TPR.
Moreover, the inter-dependence in components introduces poorly understood challenges, \eg the trade-off between minimizing the pipeline's false positives and classifying more samples statically to reduce infections.
Due to our dataset limitations, we are unable to perform such analyses.
We recommend that future research approach malware detection holistically, treating it as a pipeline, and collect suitable datasets for performing end-to-end evaluations.

Furthermore, prior work on ML-based behavioral detectors has focused strictly on a sandbox-based scenario where training and testing traces are collected from the same sandbox. 
This obscures the impact of sandbox configurations on models' robustness when deployed on traces from diverse environments, such as real-world hosts, as shown in~\cref{sec:environment_shift}.
Our work had to rely on traces from two third-party sandboxes with unknown configurations, which prevented us from conducting controlled experiments to study this impact, \eg to increase the performance in \ssbep{}.
Instead, we show that invariant learning over traces on these two sandboxes might increase robustness.
While sandbox configuration is a well-studied topic in malware analysis~\cite{liu2022enhancing,miramirkhani2017spotless,mills2020investigating}, its role in generating training sets for ML-based detectors has been overlooked.
To address this gap, we recommend that future work tackle realistic scenarios where behavioral models are deployed in unpredictable environments, emphasizing the need for robustness.
Developing principled methods for configuring sandboxes for this scenario is a promising research direction.
\section{Conclusion}
Behavioral malware detection serves as a last line of defense at endpoint hosts, providing security when all other measures have failed. 
Malware samples that breach this last line cause real-world infections and harm.
A long-standing ambition is to find the best way to detect such samples.
Through a systematic exploration of different scenarios, we provide some clarity.
State-of-the-art ML-based detectors trained on sandbox execution traces degrade severely when deployed at real-world endpoints. 
Though training on data from endpoints leads to better performance, it still falls far short of expectations established in prior evaluations.
We characterize various challenges ML approaches face in this domain and explore promising techniques targeting them, achieving moderate improvements.
Ultimately, this task remains challenging.
By exposing evaluation pitfalls and ML shortcomings, we call on the community to seek new solutions to these security-threatening problems.

\section{Acknowledgements}
This research was supported by the US Intelligence Community Postdoctoral Fellowship (Kaya), the US Department of Defense (Dumitra\c{s}), UK EPSRC Grant EP/X015971/1 (Pierazzi), US National Science Foundation Grants CNS-2154873 (Wagner) and CNS-2327427 (Botacin), and generous research awards from Google (Cavallaro and Wagner) and Amazon (Dumitra\c{s} and Wagner).
We would like to thank VirusTotal for granting us academic access to their platform.
Finally, we thank Omer Faruk Akgul and David Acs for their support in the earlier phases of our research and Erin Avllazagaj for providing us access to their program traces in the wild dataset.
Any opinions, findings, conclusions, or recommendations expressed in this material are those of the authors and do not necessarily reflect the views of the supporting organizations.

\bibliographystyle{IEEEtran}

\bibliography{references}

\begin{appendices}

\section{Trace Standardization}

\subsection{Technical Details}
\label{app:standardization}
Our EP traces contain high-level information the vendor can seamlessly collect and analyze at endpoint hosts, including file, process, and mutex creations, registry key creations and deletions, and process injections.
Sandboxes also record lower-level information (such as memory dumps~\cite{sandboxmemory}) as they are less computationally constrained.
In a standardized trace, first, we only keep the action types that all formats in our datasets share: file creation, registry key creation and deletion, process creation and injection, and mutex creation.
Second, we clean up the strings (\emph{e.g.}, file names) in each trace by removing white spaces, capitalization, punctuation, non-ASCII characters, and so on.
Third, we replace certain file and directory names to minimize differences caused by operating system versions or logging conventions.
Fourth, we use regular expressions to replace MAC addresses, Windows security and resource identifiers, hashes (SHA-256, SHA-1, MD5), and epoch timestamps with special tokens (\emph{e.g.} \texttt{<macaddress>}), to prevent introducing artifacts to our ML models.
Fifth, we tokenize the entries in each trace (\emph{e.g.}, split a full file path into its components) and save a trace as a sequence of tokens, following~\cite{jindal2019neurlux}.
We will release our sandbox dataset in this standardized format.
Moreover, this format enables researchers to submit their detectors to be evaluated against our endpoint dataset (detailed in~\cref{sec:discussion}) by providing them with an expected input format their detectors should accept.
When we train a model, we only keep the top-10K tokens in terms of frequency in the training traces and replace the remaining tokens with special tokens, such as \texttt{<rare\_file\_name>}, again following~\cite{jindal2019neurlux,trizna2023nebula}.
This makes the task more suitable for learning by eliminating uninformative tokens and reducing the feature dimensionality.

\subsection{The Performance Impact}
\label{app:standardization_impact}

We run the following experiment to understand whether our standardization hurts detection performance.
We trained \maldy models on unstandardized SB1 traces, which contain more action types (\eg file reads, registry reads, and modules loaded) than standardized traces.
We also kept the top-25K tokens (instead of the top-10K) from these traces for tokenization. 
The best model achieves 97.8\% TPR on the SB1 test set (up from 95.0\% when trained on standardized traces). 
However, its performance is 53.4\% TPR on the SB2 test set (down from 60.7\% when trained on standardized traces).
This suggests that trace standardization moderately hurts the in-domain performance (on the test traces from the training sandbox) while improving generalization to other domains (to traces from other sandboxes).
We hypothesize that standardization reduces overfitting to domain-specific features (discussed in~\cref{ssec:sandbox_artifacts}), which is desirable in our study as we aim to maximize the performance of a sandbox-based model on endpoint traces.
Note that, even after removing some features from sandbox traces, the classifier's performance is still competitive with prior work (\eg Neurlux~\cite{jindal2019neurlux}) that uses the full sandbox report.
This is because different features (\eg registry reads and registry creations) are already correlated and have redundancy, making each additional feature less effective.

\section{Further Machine Learning Details}
\label{app:ml_details}

\topic{\maldy (based on Maldy~\cite{karbab2019maldy})} We convert each trace into a list of 2-grams.
For example, the file path \texttt{a/b/c.jpg} is turned into three 2-grams: \texttt{<a/b>}, \texttt{<b/c>} and \texttt{<c.jpg>}.
As this results in an intractable number of unique 2-grams (mostly very rare), we apply the hashing trick~\cite{weinberger2009feature} that assigns a numerical value up to $2^{14}$ to each 2-gram.
The final feature vector---$x$---for a trace is a $2^{14}$-dimensional vector and each dimension is set to the number of occurrences of the corresponding 2-gram in the trace.
We use a ResNet-based architecture~\cite{gorishniy2021revisiting} to train on these features.

\topic{\neurlux (based on Neurlux~\cite{jindal2019neurlux})} We treat a trace as a natural language document and train an attention-convolution-hybrid sequence classification model.
This approach eliminates the need for feature engineering (unlike the n-gram approach) and can extract useful features from long sequences thanks to the attention mechanism.

\topic{\nebula (based on Nebula~\cite{trizna2023nebula})} We treat a trace the same way as \neurlux (a sequence) and train a self-attention transformer-based architecture that is claimed to be robust to heterogeneous information (\,  e.g., different report format).

Overall, these approaches represent an increasing level of complexity, \maldy being the simplest and most traditional and \nebula being the most advanced.
Although more advanced models seem to perform better in sandbox-based scenarios, we are interested in whether this trend changes in the endpoint scenario.

\section{How to Do Model Selection in \ssbep{}}
\label{app:model_selections}

Following our observation in~\cref{ssec:eval_sbep}, in this section, we assess whether finding a better model for \ssbep{} is possible by selecting the models using traces from an unseen sandbox (not used for training).
We rank our models (100+ in each setting) based on their SB1, SB2, and EP performances (TPR) and compute Spearman's correlation coefficients between these rankings.
Table~\ref{table:correlations} reveals that (except for one setting) \bnumber{i} the rankings based on the training sandbox correlate poorly (sometimes negatively) with EP rankings, \bnumber{ii} rankings based on an unseen sandbox correlate more strongly with EP rankings.
For example, for \nebula trained on SB1, the rankings based on SB1 and SB2 have $0.03$ and $0.45$ correlation with the rankings based on EP, respectively.
These results support our claim that it yields better results when model selection in \ssbep{} is performed using traces from a different, unseen sandbox that is not used for training.

\begin{table}[hbt]
\normalsize
\centering
\begin{adjustbox}{max width=0.95\columnwidth}
\begin{tabular}{lrrr||rrr}
\toprule
 & \multicolumn{3}{c||}{\textbf{\large Trained on SB1}} & \multicolumn{3}{c}{\textbf{\large Trained on SB2}} \\
& \multicolumn{1}{c}{SB1-SB2}  & \multicolumn{1}{c}{SB1-EP} & \multicolumn{1}{c||}{SB2-EP} &  \multicolumn{1}{c}{SB1-SB2}  & \multicolumn{1}{c}{SB1-EP} & \multicolumn{1}{c}{SB2-EP} \\
\midrule
\maldy & 0.47 & 0.03 & 0.45 & $-$0.24 & 0.39 & $-$0.46 \\ 
\neurlux & 0.88 & 0.84 & 0.75 & $-$0.17 & 0.59 & 0.33 \\ 
\nebula & 0.56 & $-$0.17 & 0.31 & 0.51 & 0.58 & 0.35 \\ 
\bottomrule
\end{tabular}
\end{adjustbox}
\caption{Ranking correlations of sandbox-trained models according to their performances (TPR) on different test sets.}
\label{table:correlations}
\vspace{-0.18in}
\end{table}

\section{Case Studies on Sandbox-Specific Artifacts}
\label{app:artifacts_dive}

Here, we dive deeper into the sandbox-specific artifacts we identified in Table~\ref{table:sandbox_artifacts}.

\textbf{\texttt{SogouExplorer}} is a Chinese web browser that exists only in malware traces (100\% MalR.) in SB1; whereas it does not exist in any SB2 traces and very few EP traces.
The samples that interact with it mainly belong to families such as \texttt{Sivis} and \texttt{Memery}, all tagged as file infectors that attach their code to other programs.
Considering that a Chinese vendor developed SB1, we believe they pre-installed this browser on their sandboxes to generate an analysis environment representative of Chinese hosts.
This, however, creates features specific to SB1 as samples interact with the programs in the environment.
Although this artifact exists in a few endpoint traces from hosts in China, its prevalence is almost zero.

\textbf{\texttt{PersonalBankPortal}}, according to our research, is a program distributed by a Chinese bank to its customers.
The samples that inject into this program belong to families such as \texttt{Tinba} and \texttt{Ramnit}, all considered as banking trojans that specifically ex-filtrate banking data.
We believe the vendor pre-installs this program to lure malware samples into exhibiting their behaviors.
Although this practice is useful for analyzing a sample~\cite{yong2021inside} (and for \ssbsb{}{}), it causes artifacts that are rarely observed in the wild.

Among the artifacts found in SB2, \textbf{\texttt{Spotify}} is a popular music streaming service, and \textbf{\texttt{Python}} is the interpreter for Python programming language.
Both programs are targeted and injected by file infectors, similar to \texttt{SogouExplorer} in SB1.
These programs are much less prevalent in endpoint traces than in SB2 traces.
We believe the SB2 vendor, based in the US, pre-installs them to create an environment representative of the hosts in the US.

\section{The Impact of Sandbox Evasion}
\label{app_sandbox_evasion}

Recent works have measured that 40-80\% of malware uses at least one evasive technique~\cite{galloro2022systematical} to avoid analysis or terminate early if it is running in a sandbox~\cite{mcafeeevasion,miramirkhani2017spotless}.
Evasion makes sandbox traces dissimilar to endpoint traces.
Although sandbox evasion is well understood, its implications for endpoint detection have not been measured.

We explore the impact of sandbox evasion.
We use a standard heuristic and treat a sample as evasive if the number of actions it performs is too low~\cite{kuchler2021does,balzarotti2010efficient,miramirkhani2017spotless}.
We first find the malware families common between our EP and SB1 test sets (30 total).
We then compute the average trace length of each common family using the traces of the samples belonging to it. 
We count only registry actions, as they can be recorded unambiguously, unlike actions such as process injections, which might have vendor-specific definitions.
We then find the length differences between SB1 and EP traces of each family.
For example, \texttt{Wannacry} and \texttt{Gandcrypt} have differences of $+11$ and $-6$, respectively.
We then split the families into two sets: the 21 families whose SB traces are longer (\emph{e.g.}, \texttt{Wannacry}) are deemed less likely to be evasive, and the 9 families whose SB traces are shorter (\emph{e.g.}, \texttt{Gandcrypt}) are deemed more likely.
The median length differences for the non-evasive and evasive families are $+4.4$ and $-4.3$, respectively, with a few outliers on both sides, \eg $+81$ for \texttt{Vobfus} and $-85$ for \texttt{Bypassuac}.
Finally, we measure the TPR of our model in \ssbsb{1}{1} and \ssbep{1} on these two malware sets individually while keeping the benign samples the same.

\begin{table}[hbt]
\normalsize
\centering
\begin{adjustbox}{max width=0.90\columnwidth}
\begin{tabular}{l|rrr|rr}
\toprule
Families & \multicolumn{1}{c}{\#Fams} & \multicolumn{1}{c}{\#EP} & \multicolumn{1}{c}{\#SB1} & \multicolumn{1}{c}{\ssbep{1}} & \multicolumn{1}{c}{\ssbsb{1}{1}} \\
\midrule
All & 30 & 169 & 5.8K & 23.1\% & 86.4\% \\
Evasive & 9 & 49 & 2.9K & 20.4\% & 89.8\% \\
Non-Eva. & 21 & 120 & 2.9K & 23.3\% & 83.0\% \\
\bottomrule
\end{tabular}
\end{adjustbox}
\caption{The performance (TPR) on malw. families more (\emph{Evasive}) or less (\emph{Non-Eva}) likely to be evading sandboxes.}
\label{table:evasion_performance_impact}
\vspace{-0.2cm}
\end{table}

The trace length of a sample monotonically increases as its execution continues.
This means the discrepancies between the execution durations in SB1 and endpoint hosts might confound our measurements.
Before presenting our results, we confirm this is unlikely to be the case: 82\% of our EP traces are from executions that lasted less than 90 seconds.
Although we do not know the exact configuration of SB1, allowing one to two minutes of execution is standard for most sandboxes in practice~\cite{kuchler2021does}.

Based on the results in Table~\ref{table:evasion_performance_impact}, we observe: \bnumber{i} \ssbep{1} performance is higher (by 3\%) on non-evasive families than on evasive families; \bnumber{ii} \ssbsb{1}{1} performance is significantly higher (by 7\%) on evasive families than on non-evasive families.
This suggests that the classifier exploits short traces (evasiveness) as a feature, which is beneficial for \ssbsb{}{}, though it does not generalize to \ssbep{}.
However, the non-trivial \ssbep{1} performance on evasive families might hint that there are still features useful for endpoint detection in the sandbox traces of evasive samples.
Although we are limited to observational data, a controlled study on evasive malware to disentangle these features is a promising direction.

\begin{figure}[H]
\centering
  \begin{subfigure}[b]{0.9\columnwidth}
    \includegraphics[width=\linewidth]{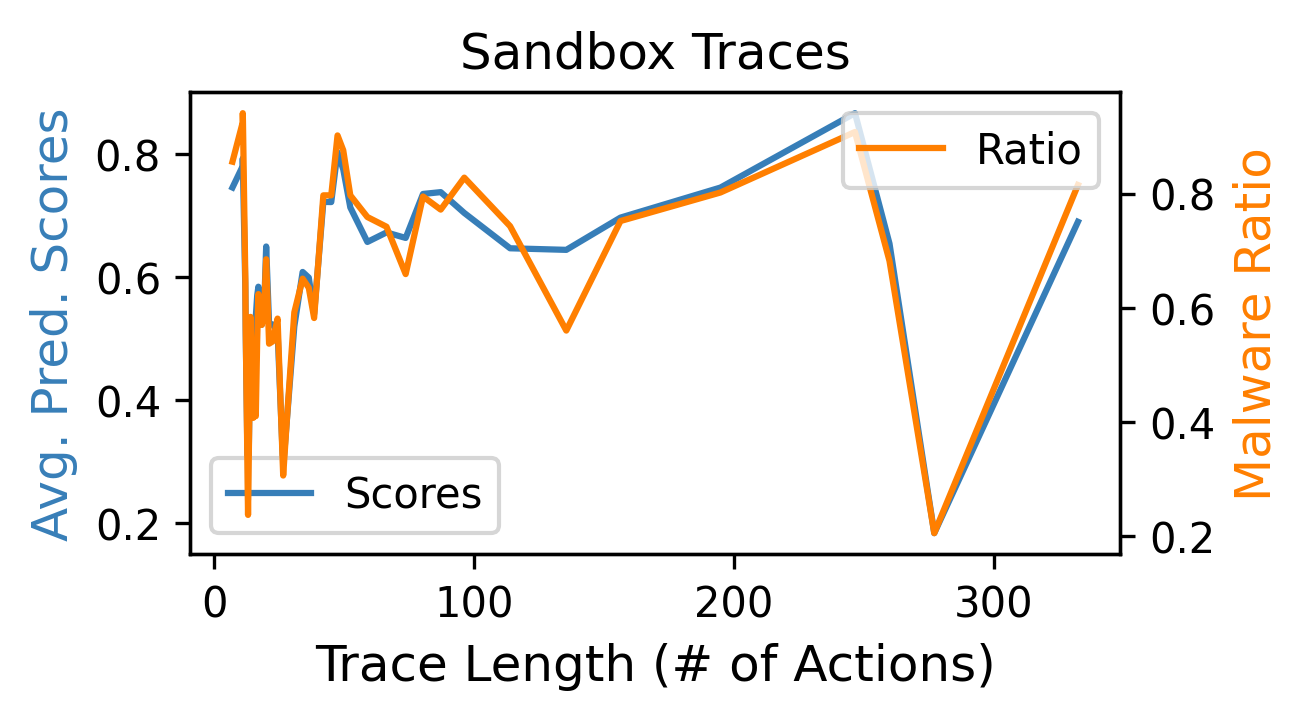}
  \end{subfigure}
  \begin{subfigure}[b]{0.9\columnwidth}
    \includegraphics[width=\linewidth]{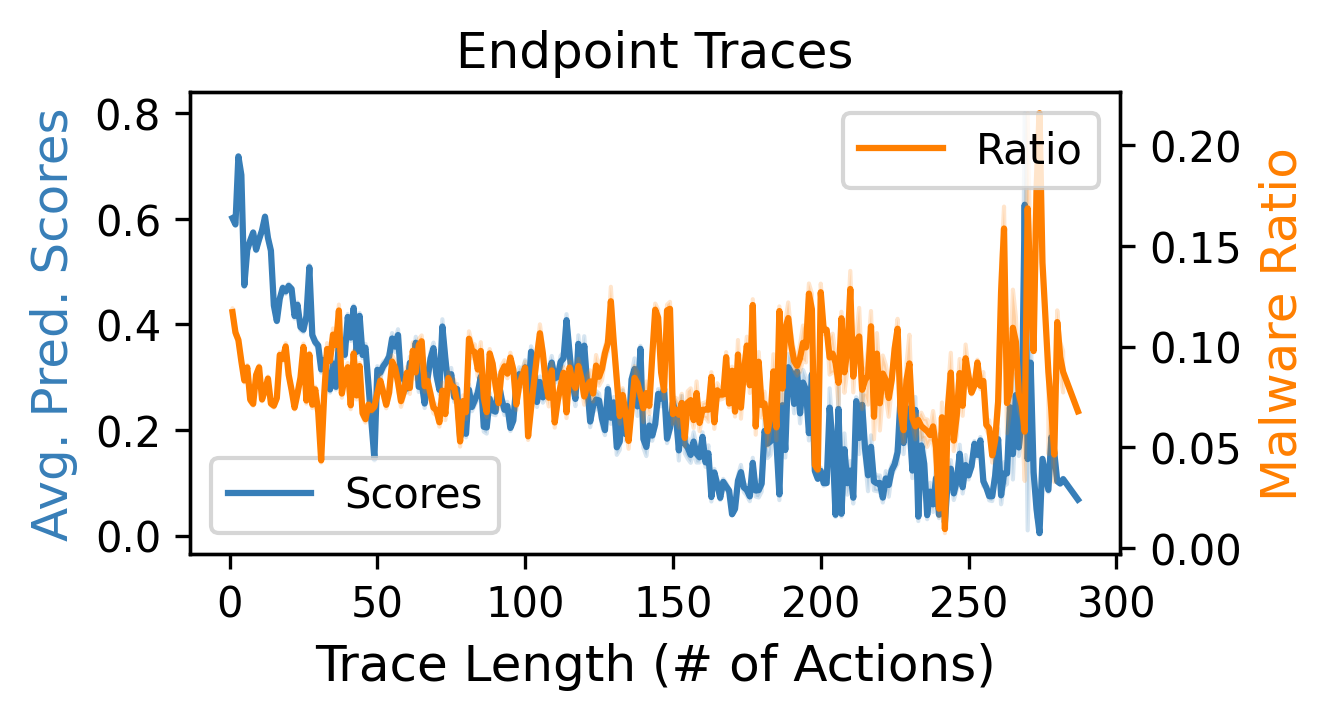}
  \end{subfigure}
  \caption{Comparing the correlations between trace length and malware-ness prediction scores of the model. \emph{Malware ratio} is the ground truth ratio of malware traces of a certain length among all traces of that length in a dataset.}
  \label{fig:length_bias}
\end{figure}

\topic{The Trace Length Bias}
Building on the previous experiment, we hypothesize that a model in \ssbep{} might learn an inverse correlation between trace length and malware-ness, \emph{i.e.}, evasive malware creates short traces, and, therefore, short traces are more likely to be malware.
In Figure~\ref{fig:length_bias}, we present the model's average predicted scores on traces with a certain length in SB1 and EP test sets.
For both SB and EP test sets, we also measure the ratio of malware traces of a given length labeled among all traces of that length.
For example, if there are 100 total traces of length $L$ and 70 of them are labeled as malware in the ground truth, the malware ratio would be 0.7 for the traces of length $L$.

These plots support our hypothesis: the model predicts a higher score for the shortest traces in both test sets, \ie it has a length bias.
This bias leads to accurate predictions in the SB set, where very short traces are much more likely to be malware (the malware ratio on the leftmost side of the upper plot is high).
However, this bias leads to inaccurate predictions in the EP set, where very short are not more likely to be malware.
On the EP traces shorter than 10 actions ($\sim$20\% of all EP traces), the model achieves 15\% TPR (vs. 20\% when all traces are kept).
Ultimately, trace length is a spurious correlation learned from sandbox traces that fails to generalize to endpoint traces.
Evidence suggests this is introduced by evasive malware that tends to produce short sandbox traces in the same execution time.
Methods preventing the model from learning such correlations~\cite{kirichenko2023last,pezeshki2021gradient} offer a promising next step.

\section{Fine-Tuning a Sandbox-Based Model on Endpoint Traces}
\label{app:fine_tuning}
\begin{figure}[H]
\centering
\includegraphics[width=\linewidth]{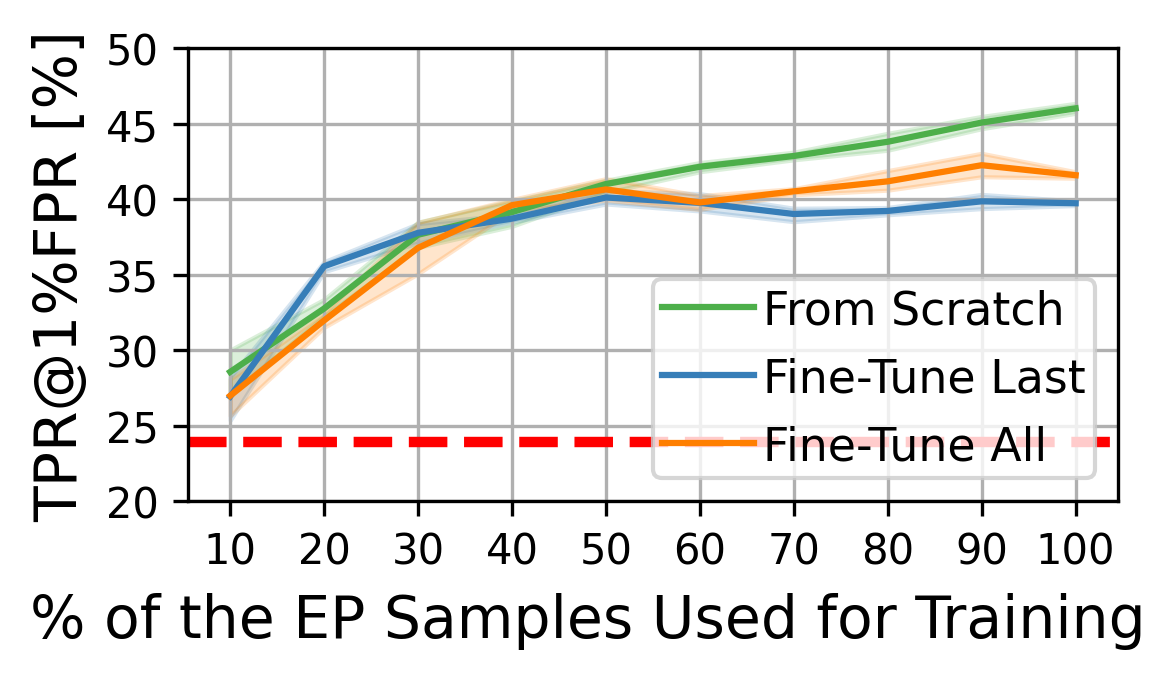}\par\medskip
  \caption{The results of fine-tuning a sandbox-based model in \ssbep{} on endpoint traces. The dashed line indicates our best model  \ssbep{}. Experiments on \maldy.}
  \label{fig:fine_tuning_sb}
\end{figure}

For the experiments in Figure~\ref{fig:fine_tuning_sb}, we select two endpoint traces per sample and implement two fine-tuning strategies: \textbf{(i)} freezing the encoder layers $enc$ of our model and tuning only the classification layer $g$, and \textbf{(ii)} tuning all layers without freezing.
We train models on increasing portions of the samples in our EP training set (randomly selected) and average the results over 10 models.
We make the following observations.
In low-data regimes (below 30\% of the EP data), fine-tuning only $g$ outperforms the other options in terms of TPR by $\sim$3--5\%.
However, with more data, it starts to perform significantly worse, due to being less flexible in learning from the EP traces.
Finally, fine-tuning all layers is generally the worst option, and training from scratch is the best when more EP data is available.
These experiments show that starting from a well-performing model in \ssbep{} is beneficial in regimes with limited EP data, highlighting a promising direction for future work.

\begin{table*}[hbt]
\normalsize
\begin{adjustbox}{max width=\textwidth}
\begin{tabular}{l|rrr|rrr|rrr|rrr|rrr|rrr|rrr}
\toprule
\multirow{3}{*}{\shortstack[l]{{Sel.}\\{Test}\\{Set}}} & 
\multicolumn{9}{c|}{\textbf{\large Trained on SB1}} & 
\multicolumn{9}{c|}{\textbf{\large Trained on SB2}} & 
\multicolumn{3}{c}{\textbf{\large Trained on EP}} \\

& \multicolumn{3}{c|}{\normalsize \ssbsb{1}{1}} &  
\multicolumn{3}{c|}{\normalsize \ssbsb{1}{2}} & 
\multicolumn{3}{c|}{\normalsize \ssbep{1}} & 
\multicolumn{3}{c|}{\normalsize \ssbsb{2}{1}} & 
\multicolumn{3}{c|}{\normalsize \ssbsb{2}{2}} & 
\multicolumn{3}{c|}{\normalsize \ssbep{2}} &
\multicolumn{3}{c}{\normalsize \sepep} \\ 

 & \maldy & \neurlux & \nebula & \maldy & \neurlux & \nebula & \maldy & \neurlux & \nebula & \maldy & \neurlux & \nebula & \maldy & \neurlux & \nebula & \maldy & \neurlux & \nebula & \maldy & \neurlux & \nebula \\
SB1 & 
99.1 & 99.0 & 98.9 & 
93.8 & 91.3 & 92.6 & 
76.1 & 70.7 & 72.8 & 

89.5 & 83.4 & 89.1 & 
97.7 & 94.1 & 96.9 & 
72.0 & 68.2 & 70.8 & 

85.6 & 85.2 & 84.3 \\  

SB2 &  
99.0 & 98.6 & 98.8 & 
94.9 & 93.2 & 93.5 & 
76.6 & 71.3 & 74.4 & 

81.7 & 71.5 & 86.4 & 
98.7 & 98.5 & 98.2 & 
71.6 & 66.7 & 72.7 & 

85.4 & 85.9 & 84.8 \\  

EP &  

99.0 & 98.6 & 98.7 & 
92.9 & 91.2 & 92.9 & 
\color{red}{\textbf{78.4}} & 74.1 & 74.9 & 

84.4 & 71.8 & 88.3 & 
98.0 & 97.8 & 98.0 & 
74.6 & 72.0 & 73.4 & 

\color{red}{\textbf{87.5}} & 86.8 & 86.8 \\  

\bottomrule
\end{tabular}
\end{adjustbox}
\caption{The performance (AUC\%) of three ML approaches (\maldy, \neurlux, \nebula) in seven detection scenarios (based on Table~\ref{table:detection_scenarios}). The models are selected using the test set in each row, and the average AUC of the top 20 models is reported.}
\label{table:app_auc_results}
\end{table*}

\section{More Details on the Artifact Release}
\label{app:leaderboard}

We will release the following artifacts to the community (visit our website for more details \href{https://malwaredetectioninthewild.github.io/}{https://malwaredetectioninthewild.github.io/}):

\topic{Sandbox Dataset}
We will release the training portion of our sandbox datasets in Table~\ref{table:dataset_stats}, stored in the standardized format discussed in~\cref{ssec:datasets}. 
We avoid releasing the testing portion publicly to prevent researchers who wish to participate in our realistic evaluation leaderboard from obtaining an impractical advantage by training their detectors on it (will be released if requested in exchange for being excluded from the leaderboard).
This will level the playing field for participants and ensure they all have access to the same sandbox data for development.
Further, as this data contains traces from two sandboxes, participants can leverage our observations in~\cref{sec:sb_limitations} and tune their hyper-parameters on a second sandbox, not seen during training, for improving the endpoint performance or train their models on data from both sandboxes.
They can also apply invariant learning techniques, which have shown promising results in~\cref{ssec:env_invariance}.

\topic{Sample Metadata}
We will release the metadata relating to the training samples in all datasets, including their SHA-256 hashes, ground truth labels, family tags (if malware), and first-seen timestamps.
Additionally, our metadata also annotates the source of a sample, \emph{e.g.}, EMBER~\cite{anderson2018ember}, SOREL~\cite{harang2020sorel20m}, or our endpoint dataset.
This allows researchers to compute realistic priors over malware families for endpoint detection and make use of our findings and improvements in~\cref{ssec:resampling}.
Moreover, they can also create realistic testing distributions over samples using these malware family priors (and avoid the problems we discussed in~\cref{sec:sample_shift}.

\section{Additional Tables and Figures}

\begin{figure}[H]
\centering
\ndtopic{Statistics on the Endpoint Traces}\par\medskip
  \begin{subfigure}[b]{0.48\columnwidth}
    \includegraphics[width=\linewidth]{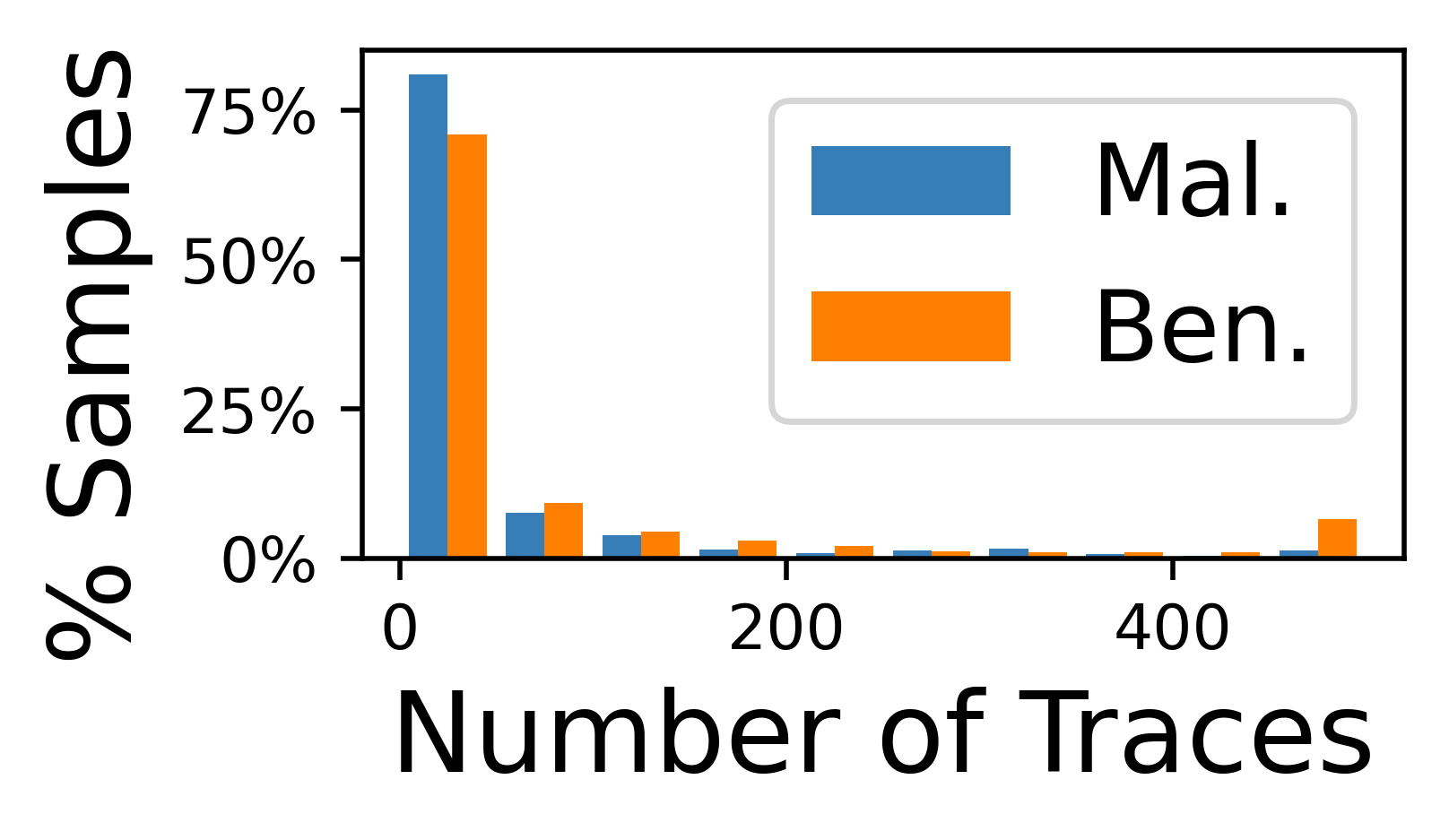}
    \caption{The number of traces per sample.}
    \label{fig:trace_stats_numbers}
  \end{subfigure}\hfill
  \begin{subfigure}[b]{0.48\columnwidth}
    \includegraphics[width=\linewidth]{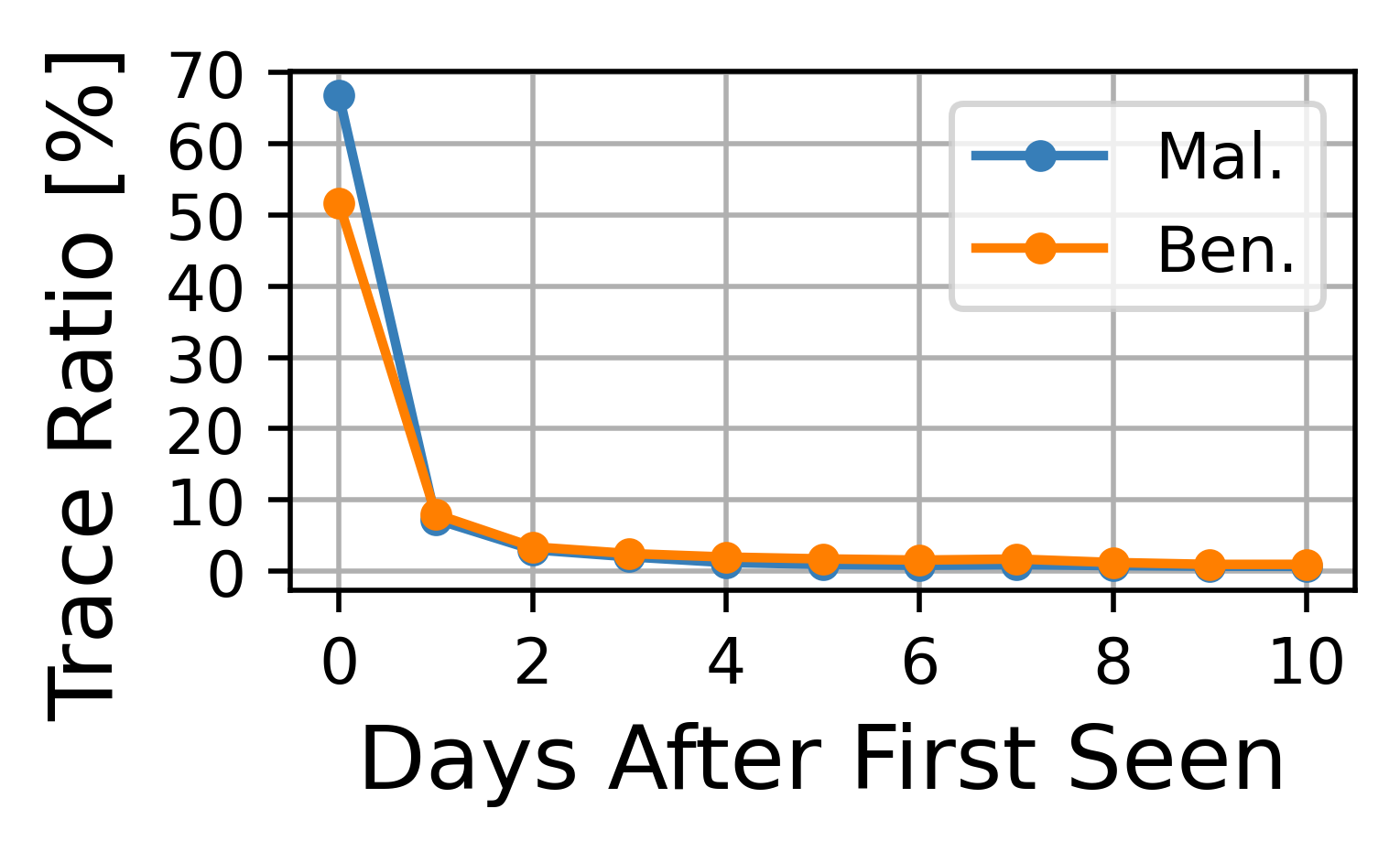}
    \caption{The average ratio of seen traces per sample as a function of time.}
    \label{fig:trace_stats_days}
  \end{subfigure}
  \caption{Statistics on the endpoint traces in our dataset.}
  \label{fig:trace_stats}
\end{figure}

\begin{figure}[htb]
\ndtopic{Soft Labeling Function}
\centering 
\includegraphics[width=0.9\columnwidth]{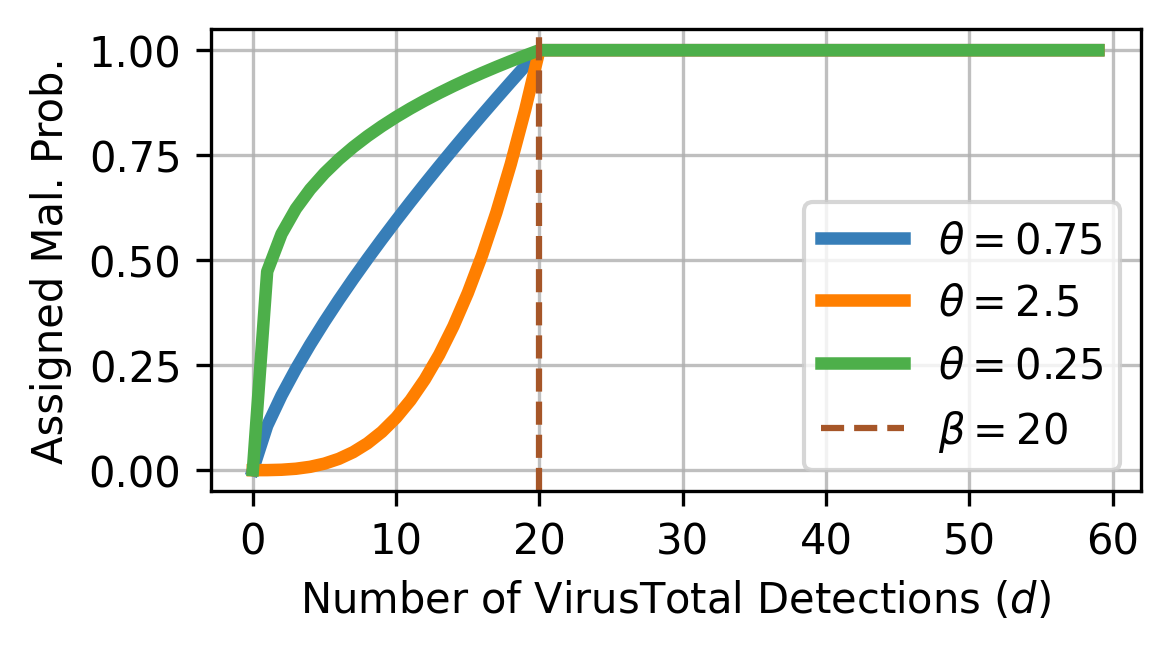}\par\medskip
\caption{Our function for assigning probabilistic (soft) labels to each sample based on the number of VirusTotal detections.}
\label{fig:soft_label_func}
\end{figure}

\begin{figure}[H]
\centering
    \ndtopic{Histogram of Prediction Score Standard Deviations on Endpoint Traces of the Same Sample for Models Trained With the Invariance Loss.}\par\medskip
  \begin{subfigure}[b]{0.48\columnwidth}
    \includegraphics[width=\linewidth]{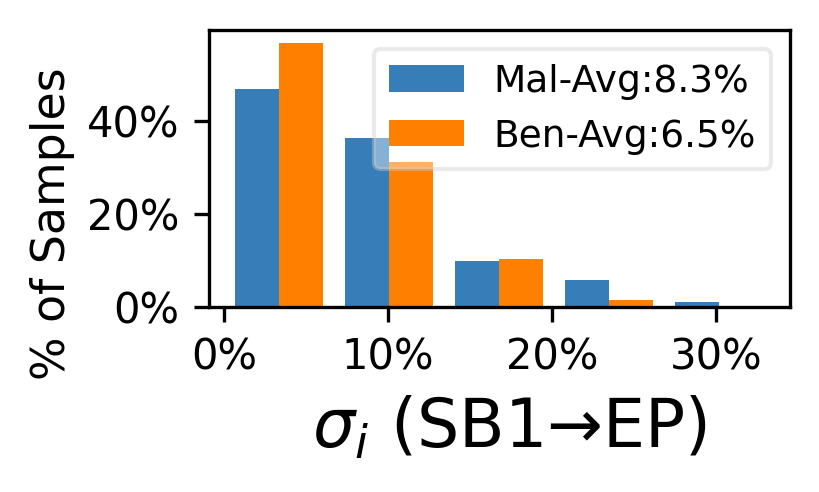}
  \end{subfigure}
  \begin{subfigure}[b]{0.48\columnwidth}
    \includegraphics[width=\linewidth]{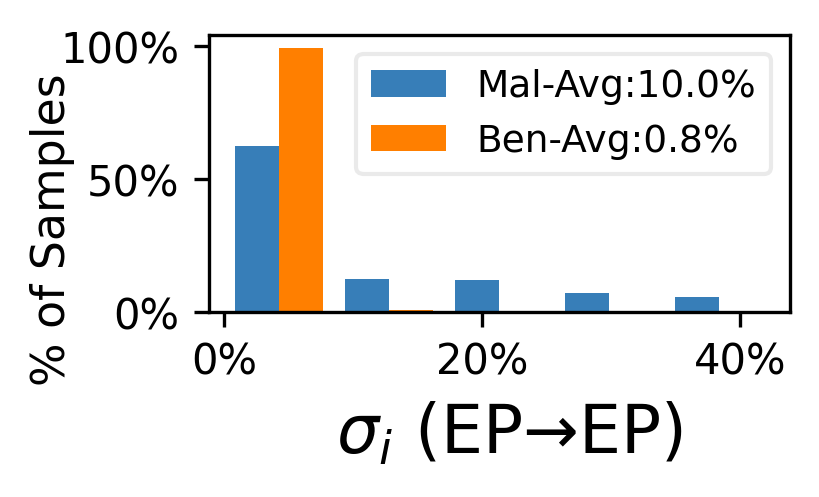}
  \end{subfigure}
\caption{Prediction score standard deviations on the endpoint traces of each sample in the EP test set; for a model in \ssbep{1} (\emph{left}), and \sepep \emph{(right)}.} 
    \label{fig:ep_std_devs_w_env}
\end{figure}

\begin{table}[H]
\normalsize
\centering
\ndtopic{Top Malware Families and Benign Publishers}\par\medskip
\begin{adjustbox}{max width=\columnwidth}
\begin{tabular}{lrr||lrr}
\toprule
\multicolumn{3}{c||}{\textbf{\small{EP Test Set}}} & \multicolumn{3}{c}{\textbf{\small{SB Test Set}}} \\ 
Name & EP\% & SB1\% & Name & SB1\% & EP\%  \\ 
\midrule
\texttt{GENERIC} & 23.8\% & 4.8\% & \texttt{GENERIC} & 4.8\% & 23.8\%  \\
\texttt{Chindo} & 9.0\% & 0.0\% & \texttt{Gepys} & 4.4\% & 0.0\% \\
\texttt{Emotet} & 7.6\% & 0.5\% & \texttt{Sivis} & 4.3\% & 0.3\%   \\
\texttt{Gandcrab} & 6.0\% & 3.6\% & \texttt{Flystudio} & 4.2\% & 0.3\%  \\
\texttt{Loadmoney} & 6.0\% & 0.1\% & \texttt{Upatre} & 3.8\% & 0.8\% \\
\texttt{Khalesi} & 5.0\% & 0.3\% & \texttt{Gandcrab} & 3.6\% & 6.0\% \\
\texttt{Installcore} & 4.5\% & 0.0\% & \texttt{Shipup} & 3.5\% & 0.0\% \\
\midrule
\midrule
\texttt{UNSIGNED} & 29.6\% & 47.2\% & \texttt{UNSIGNED} & 47.2\% & 28.4\% \\
\texttt{Microsoft} & 7.8\% & 0.4\% & \texttt{Google} & 4.7\% & 0.2\% \\
\texttt{Tencent} & 2.6\% & 0.5\% & \texttt{Mozilla} & 3.7\% & 0.1\% \\
\texttt{Qihoo} & 2.2\% & 0.2\% & \texttt{Digital R.} & 3.5\% & 0.0\% \\
\texttt{Zoho} & 1.3\% & 0.1\% & \texttt{Yandex} & 2.7\% & 1.0\% \\
\texttt{Opera} & 1.2\% & 0.1\% & \texttt{ScreenC.} & 2.2\% & 0.0\% \\
\texttt{Yandex} & 1.0\% & 2.7\% & \texttt{Zoom} & 2.0\% & 0.3\% \\
\bottomrule
\end{tabular}
\end{adjustbox}
\caption{Top malware families (\emph{top}) and benign (\emph{bottom}) publishers in our EP (\emph{left}) and SB (\emph{right}) test sets, along with their shares in each dataset.}
\label{table:top_fams_pubs}
\vspace{-0.2cm}
\end{table}

\begin{table}[hbt]
\large
\centering
\ndtopic{Case Study on \texttt{Wannacry} and Its Indicators-Of-Compromise}\par\medskip

\begin{adjustbox}{max width=0.9\columnwidth}
\begin{tabular}{l|rrr|rrr}
\toprule
\multirow{2}{*}{\shortstack[l]{Trace \\ Type}} & \multicolumn{3}{c|}{\textbf{Ratio}} & \multicolumn{3}{c}{\textbf{Avg. Pred. Score}} \\ 
& \multicolumn{1}{c}{SB1} & \multicolumn{1}{c}{SB2} & \multicolumn{1}{c|}{EP} & \multicolumn{1}{c}{\ssbep{1}} & \multicolumn{1}{c}{\ssbep{2}} & \multicolumn{1}{c}{\sepep} \\
\midrule
Any IOC & 89.9\% & 100\% & 97.7\% & 82.0\% & 75.6\% & 97.4\% \\
No IOC & 10.1\% & 0.0\% & 2.3\% & 77.7\% & 67.7\% & 51.8\%  \\
\bottomrule
\end{tabular}
\end{adjustbox}
\caption{For \texttt{Wannacry} traces in our SB1, SB2, and EP datasets, we first measure the ratio of traces with at least one and no IOC. We then measure the average prediction score of our \maldy models in \ssbep{1}, \ssbep{2} and \sepep on these traces.}
\label{table:wannacry_iocs}
\end{table}

\end{appendices}

\end{document}